\documentclass[reqno,12pt,dvips]{amsart}

\usepackage{amssymb,times,mathptmx,cite,eucal,psfrag,array,setspace,geometry}
\usepackage[dvips]{graphicx}

\numberwithin{equation}{section}

\allowdisplaybreaks

\newcolumntype{C}{>{$}c<{$}} 

\newcommand{\group}[1]{\mathsf{#1}}
\newcommand{\alg}[1]{\mathfrak{#1}}

\newcommand{\uealg}[1]{\mathcal{U} \bigl( #1 \bigr)}

\newcommand{\func}[2]{#1 \left( #2 \right)}

\newcommand{\brac}[1]{\left( #1 \right)}
\newcommand{\sqbrac}[1]{\left[ #1 \right]}

\newcommand{\set}[1]{\left\{ #1 \right\}}

\newcommand{\abs}[1]{\left| #1 \right|}

\newcommand{\ZZ}{\mathbb{Z}}

\newcommand{\RR}{\mathbb{R}}
\newcommand{\CC}{\mathbb{C}}

\newcommand{\dd}{\mathrm{d}}
\newcommand{\ii}{\mathfrak{i}}

\newcommand{\eps}{\varepsilon}

\newcommand{\comm}[2]{\bigl[ #1 , #2 \bigr]}
\newcommand{\acomm}[2]{\bigl\{ #1 , #2 \bigr\}}

\newcommand{\bra}[1]{\bigl\langle #1 \bigr\rvert}
\newcommand{\ket}[1]{\bigl\lvert #1 \bigr\rangle}
\newcommand{\braket}[2]{\bigl\langle #1 \bigr\rvert \bigl. #2 \bigr\rangle}
\newcommand{\bracket}[3]{\bigl\langle #1 \bigr\rvert #2 \bigl\lvert #3 \bigr\rangle} 

\newcommand{\radord}[1]{#1}
\newcommand{\normord}[1]{{} : #1 : {}} 

\newcommand{\MinMod}[2]{\mathcal{M} \left( #1 , #2 \right)}
\newcommand{\sMinMod}[2]{\mathcal{SM} \left( #1 , #2 \right)}

\newcommand{\VirVerMod}[2]{\widetilde{\mathcal{V}}^{#1}_{#2}}
\newcommand{\VirIrrMod}[2]{\widetilde{\mathcal{L}}^{#1}_{#2}}
\newcommand{\ExtVerMod}[2]{\mathcal{V}^{#1}_{#2}}
\newcommand{\ExtIrrMod}[2]{\mathcal{L}^{#1}_{#2}}

\newcommand{\gcreq}[1]{\overset{ #1 }{=}}

\newcommand{\eqnref}[1]{Equation~(\ref{#1})}
\newcommand{\eqnDref}[2]{Equations~(\ref{#1}) and (\ref{#2})}
\newcommand{\eqnsref}[2]{Equations~(\ref{#1}--\ref{#2})}
\newcommand{\appref}[1]{Appendix~\ref{#1}}
\newcommand{\secref}[1]{Section~\ref{#1}}
\newcommand{\figref}[1]{Figure~\ref{#1}}
\newcommand{\tabref}[1]{Table~\ref{#1}}
\newcommand{\propref}[1]{Proposition~\ref{#1}}

\newcommand{\thmref}[1]{Theorem~\ref{#1}}
\newcommand{\corref}[1]{Corollary~\ref{#1}}
\newcommand{\correfs}[2]{Corollaries~\ref{#1} and \ref{#2}}

\newcommand{\cft}{conformal field theory}
\newcommand{\cfts}{conformal field theories}
\newcommand{\ope}{operator product expansion}
\newcommand{\opes}{operator product expansions}
\newcommand{\hws}{highest weight state}
\newcommand{\hwss}{highest weight states}

\newcommand{\hwms}{highest weight modules}
\newcommand{\uea}{universal enveloping algebra}
\newcommand{\gcr}{generalised commutation relation}
\newcommand{\gcrs}{generalised commutation relations}

\DeclareMathOperator{\id}{id}

\newtheorem{theorem}{Theorem}[section]				
\newtheorem{proposition}[theorem]{Proposition}	
\newtheorem{corollary}[theorem]{Corollary}		
\newtheorem*{claim}{Claim}

\begin{document}

\title[Minimal Model Extended Algebras]{The Extended Algebra of the Minimal Models}

\author[P Mathieu]{Pierre Mathieu}

\address[Pierre Mathieu]{
D\'{e}partement de Physique, de G\'{e}nie Physique et d'Optique \\
Universit\'{e} Laval \\
Qu\'{e}bec, Canada G1K 7P4
}

\email{pmathieu@phy.ulaval.ca}

\author[D Ridout]{David Ridout}

\address[David Ridout]{
D\'{e}partement de Physique, de G\'{e}nie Physique et d'Optique \\
Universit\'{e} Laval \\
Qu\'{e}bec, Canada G1K 7P4
}

\email{darid@phy.ulaval.ca}

\thanks{\today}

\begin{abstract}
The minimal models $\MinMod{p'}{p}$ with $p'>2$ have a unique (non-trivial) simple current of conformal dimension $h = \tfrac{1}{4} \brac{p' - 2} \brac{p - 2}$.  The representation theory of the extended algebra defined by this simple current is investigated in detail.  All highest weight representations are proved to be irreducible:  There are thus no singular vectors in the extended theory.  This has interesting structural consequences.  In particular, it leads to a recursive method for computing the various terms appearing in the \ope{} of the simple current with itself.  The simplest extended models are analysed in detail and the question of equivalence of \cfts{} is carefully examined.
\end{abstract}

\maketitle

\onehalfspacing

\section{Introduction} \label{secIntro}


This article is the second in a series in which we analyse well-known \cfts{} from the viewpoint of an extended symmetry algebra defined by a simple current.  Here, we present the minimal model case.  This series was initiated by \cite{RidSU206}, in which we constructed and studied the extended algebras of the $\func{\group{SU}}{2}$ Wess-Zumino-Witten models.  Whilst these two articles are formally independent, we will see that the main results obtained in \cite{RidSU206} have formal analogues in the study of the minimal model extended algebras.  The construction of the latter is in fact easier in many respects, because the algebra is only graded by the conformal dimension, and not additionally by the $\func{\alg{sl}}{2}$-weight.  However, this structural reduction makes the analysis of the minimal model extended algebras somewhat more difficult than the $\func{\group{SU}}{2}$ case.

\subsection{Motivation} \label{secMotiv}

The reformulation of the minimal models from the point of view of the extended algebra generated by their simple current fits within the general program of trying to understand and/or derive fermionic character formulae (see for example \cite{KedSum95}) by intrinsic conformal field theoretical methods.  Such fermionic formulae reflect the description of the space of states in terms of quasi-particles subject to some restriction rules.  Within the framework developed here, the quasi-particles are represented by the modes of the simple current.  The first objective towards constructing the fermionic formula is to derive a complete set of constraints on strings of these simple current modes so as to obtain a complete description of the space of states.  The character will then be the generating functions of these states.

The best studied  minimal models are those with $p'=2$.  Their basis of states has been derived in \cite{FeiAnn92} (through an analysis inspired in part by that of the parafermionic models in \cite{LepStr85}).  For fixed $p$, the basis is formulated in terms of the Virasoro modes and it is controlled by partitions with difference $2$ at distance $k$, where $p=2k+3$.  However, these models do not fit within our framework as they possess no simple current.  More precisely, the simple current of the $\MinMod{2}{p}$ models reduces to the identity field.

The next simplest cases are the $\MinMod{3}{p}$ models.  Here, there exists a non-trivial simple current, namely $\phi_{2,1}$.  These minimal models have been reformulated from the point of view of the extended algebra generated by $\phi_{2,1}$ in \cite{JacQua06}, and this reformulation has led to the proposal of a basis of states expressed solely in terms of the $\phi_{2,1}$ modes\footnote{Note that another proposed basis has been presented in \cite{JacEmb06}, without much rationale.  This alternative basis involves both the Virasoro and $\phi_{2,1}$ modes and it unifies, in a sense, the $p'=2$ and $p'=3$ bases.} (see also related results in \cite{FeiVer02}).  Although some justifications for the linear independence and spanning property of the states were discussed, no formal proofs were presented, although the resulting fermionic characters were derived, reproducing known expressions (given for instance in \cite{BytFer99}).

There are two natural ways to extend these results to $p'>3$.  One can either pursue the construction of a basis in terms of $\phi_{2,1}$ --- noting that for $p'>3$, $\phi_{2,1}$ is no longer a simple current --- or look for an extended formulation in terms of the new model's simple current.  The description of the basis of states for the $\MinMod{p'}{p}$ models with $p'>3$ in terms of the $\phi_{2,1}$ modes is plagued with serious technical difficulties rooted in the fact that the \ope{} of $\phi_{2,1}$ with itself has more than one channel.  This construction is pursued in \cite{FeiMon05}.  Here we follow the second approach, which amounts to giving the role of basic generator, played by $\phi_{2,1}$ in the $\MinMod{3}{p}$ extended algebra, to the corresponding simple current of the general $\MinMod{p'}{p}$ model, $p' \geqslant 3$.  This simple current is $\phi_{p'-1,1}$ (\secref{secNotation}).

In \cite{JacQua06}, the $\MinMod{3}{p}$ basis of states is formulated in terms of $p$-dependent constraints on the strings of modes of $\phi_{2,1}$ at distance 1 and 2.  However, the precise mechanism by which these constraints arise was not clearly isolated\footnote{See in particular the discussion presented in the two paragraphs before \cite[Eqn.~(3.16)]{JacQua06}.}.  Do these conditions on the basis of states follow solely from the \ope{} of $\phi_{1,2}$ with itself, by considering a sufficiently large number of terms (as exploratory computations seem to indicate)?  Do the vacuum and simple current singular vectors play any role with regard to these constraints?  Moreover, in   \cite{JacQua06} it was implicitly assumed that there are no singular vectors in the extended algebra, so that no subtractions were needed to construct the character.  However, no arguments supporting this assumption were presented there, apart from the fact that ignoring such potential singular vectors did indeed reproduce the correct characters.  Finally, all issues relating to linear independence of the proposed basis could be treated rather informally because the fermionic characters, with which the state generating function could be compared, were already known in precisely the same form.  

For $p'>3$, the situation is completely different.  Although fermionic expressions are known for all irreducible minimal models \cite{WelFer05}, they are not obviously of the type that would suit the combinatorial descriptions of the bases found for $p'=3$.  To make progress then, we require definite conclusions concerning the representation theory of the algebra defined by the simple current modes.   It is therefore mandatory (both for the $p'=3$ case and its generalisation) to consolidate the foundation and reanalyse carefully the mathematical formalism implied by this simple current approach.  This is the subject of the present paper.  Basis issues will be reported in a sequel \cite{RidMin07}.

We mention that reformulating theories through a quasi-particle approach has many physical advantages and applications.  As some of these have already been discussed in \cite{RidSU206}, we will not address them further here.

\subsection{Outline} \label{secOutline}

The article is organised as follows.  Some preliminary results are first collected in the following subsection, where we also fix the notation.  The extended algebra itself is defined in \secref{secStructure}.  In short, this algebra is defined by the \ope{} of the simple current $\phi_{p'-1,1}$ with itself.  We first examine the commutativity and associativity of the defining \ope{}, and a neat (rigorous) argument is presented for the necessity of introducing an $\mathcal{S}$-type operator in this expansion, anticommuting with the simple current modes when $4 h_{p'-1,1}$ is odd.  Whilst the necessity of such an operator is not seen at the level of the full (non-chiral) theory (which explains why it was not observed until recently), we show that its presence is nevertheless essential to a consistent treatment of the representation theory.

It is then a simple matter to extend the Virasoro algebra to the algebra defined by the simple current modes.  An important consequence of the dimension of $\phi_{p'-1,1}$ not necessarily being integral or half-integral is that the defining relations of the algebra of modes must be formulated in terms of \gcrs{} involving infinite sums.  The representation theory of this \emph{extended algebra} is then developed in \secref{secRepTheory}.  We emphasise here the central role played by the monodromy charge.  Symmetry properties of the \gcrs{} are also studied, and we identify for later use those \gcrs{}, amongst the many that we generate, which are actually independent.

Various examples of extended minimal models are presented in \secref{secCFTEquivs}.  In the context of extended algebras, we clarify the meaning of equivalence of two conformal field theories.  We do this by introducing a sequence of (partially) extended models whose infinite limit is the genuine extension under study.  These ``incomplete'' versions are in essence defined by truncating the depth at which the \ope{} of $\phi_{p'-1,1}$ with itself is probed. 

The central result of the paper is then presented in \secref{secSingVect}.  There it is argued that in the extended algebra, all modules are free of singular vectors.  This is first supported by a number of explicit computations within the simplest minimal models.  Then, a generic argument is presented which demonstrates that in the vacuum module of the extended theory, there are no singular vectors.  In other words, the vacuum Verma module is already irreducible.  Finally, an appeal to a (claimed) general property of the Virasoro vacuum singular vector (the one at grade $\brac{p'-1} \brac{p-1}$) is used to lift this conclusion to the other (extended) highest weight modules appearing in the theory.  The claimed general property in this case is that the corresponding null field completely controls the spectrum of the minimal model as well as \emph{all} of its singular vectors.

Restated differently, our main result states that when taking into account sufficiently many of the terms in the \ope{} of $\phi_{p'-1,1}$ with itself, the vacuum and the simple current principal singular vectors identically vanish (and we claim that this then implies the vanishing of the other singular vectors).  This means that the corresponding Virasoro singular vectors are coded as identities in the model's defining \ope{}.  But once this is established, it can be turned around and used to deduce recurrence relations for the various terms of this \ope{}.  Such relations are derived in \secref{secApps}.  These appear to be rather powerful technical tools, as the displayed illustrative computations exemplify.

\subsection{Notation and Preliminaries} \label{secNotation}

The minimal models may be defined as those two-dimensional \cfts{} whose state space is constructed out of a finite number of irreducible representations of the Virasoro algebra $\alg{Vir}$.  This is the infinite dimensional Lie algebra spanned by the modes $L_n$ ($n \in \ZZ$) and $C$, subject to the commutation relations
\begin{equation} \label{eqnVirComm}
\comm{L_m}{L_n} = \brac{m - n} L_{m + n} + \binom{m + 1}{3} \delta_{m + n , 0}  \frac{C}{2} \qquad \text{and} \qquad \comm{L_m}{C} = 0.
\end{equation}
The $L_n$ are the modes of the energy-momentum field
\begin{equation}
\func{T}{z} = \sum_n L_n z^{-n-2},
\end{equation}
whose \ope{} is
\begin{equation} \label{eqnOPETT}
\radord{\func{T}{z} \func{T}{w}} = \frac{C/2}{\brac{z-w}^4} + \frac{2 \func{T}{w}}{\brac{z-w}^2} + \frac{\func{\partial T}{w}}{z-w} + \ldots
\end{equation}
The generator $C$ is central, and acts on the irreducible representations comprising the theory as multiplication by $c$, the central charge.

Up to a finite ambiguity in the modular invariant, the minimal models are classified by two coprime integers $p > p' > 1$ (in fact, by their ratio).  We will generally denote a minimal model by $\MinMod{p'}{p}$, presuming that the modular invariant is diagonal unless otherwise specified (we will actually only be concerned with the structure of the theory at the chiral level).  The central charge of $\MinMod{p'}{p}$ is given by
\begin{equation}
c = 1 - \frac{6 \brac{p - p'}^2}{p p'},
\end{equation}
and the irreducible $\alg{Vir}$-modules which appear (the spectrum) have \hwss{} of conformal dimension
\begin{equation} \label{eqnConfDim}
h_{r,s} = \frac{\brac{p r - p' s}^2 - \brac{p - p'}^2}{4 p p'},
\end{equation}
where $1 \leqslant r \leqslant p'-1$ and $1 \leqslant s \leqslant p-1$.  We note the symmetry $h_{p'-r,p-s} = h_{r,s}$ which implies a redundancy in the spectrum.  We will generally denote a primary field of conformal dimension $h_{r,s}$ by $\func{\phi_{r,s}}{z}$, and the corresponding \hws{} by $\ket{\phi_{r,s}}$.  The aforementioned symmetry is then expressed by the field identifications $\func{\phi_{r,s}}{z} = \func{\phi_{p'-r,p-s}}{z}$.  In particular, the identity field is $\func{\phi_{1,1}}{z} = \func{\phi_{p'-1,p-1}}{z}$, and the corresponding \hws{} is the vacuum, which we denote by $\ket{0} \equiv \ket{\phi_{1,1}}$.

The fusion rules for the (diagonal) minimal models are known explicitly \cite{BelInf84,DiFCon97}, and read
\begin{equation}
\phi_{r,s} \times \phi_{r',s'} = \mspace{10mu} \sideset{}{'} \sum_{m = 1 + \abs{r - r'}}^{\min \set{r + r' - 1 , 2p' - r - r' - 1}} \mspace{10mu} \sideset{}{'} \sum_{n = 1 + \abs{s - s'}}^{\min \set{s + s' - 1 , 2p - s - s' - 1}} \phi_{m,n},
\end{equation}
where the primed summation indicates that $m$ and $n$ increment in twos.  We are interested in so-called \emph{simple currents}, fields whose fusion with any primary field gives back a single primary field.  We see that $\phi_{r',s'}$ is therefore a simple current precisely when 
\begin{align}
1 + \abs{r - r'} &= \min \set{r + r' - 1 , 2p' - r - r' - 1} & &\text{for all } 1 \leqslant r \leqslant p'-1, \label{eqnSimpCurr1} \\
\text{and} \qquad 1 + \abs{s - s'} &= \min \set{s + s' - 1 , 2p - s - s' - 1} & &\text{for all } 1 \leqslant s \leqslant p-1. \label{eqnSimpCurr2}
\end{align}
We consider \eqnref{eqnSimpCurr2} first.  When $s + s' \leqslant p$, this becomes $\abs{s - s'} = s + s' - 2$, so squaring gives
\begin{equation}
\brac{s - 1} \brac{s' - 1} = 0 \qquad \text{for all $1 \leqslant s \leqslant p - s'$.}
\end{equation}
Thus $s' = 1$ or $s' = p-1$ (forcing $s = 1$).  Similarly, when $s + s' \geqslant p$, $\abs{s - s'} = 2p - s - s' - 2$, so
\begin{equation}
\brac{s - p + 1} \brac{s' - p + 1} = 0 \qquad \text{for all $p - s' \leqslant s \leqslant p - 1$.}
\end{equation}
Again, $s' = p-1$ or $s' = 1$ (forcing $s = p-1$).  The analysis of \eqnref{eqnSimpCurr1} is identical, so we find that the only simple currents are
\begin{equation}
\phi_{1,1} = \phi_{p'-1,p-1} \qquad \text{and} \qquad \phi_{1,p-1} = \phi_{p'-1,1}.
\end{equation}
For $p' = 2$, these coincide, so there are no non-trivial simple currents.  For $p' > 2$, there is a unique non-trivial simple current $\phi_{p'-1,1}$, which we will generally denote by $\phi$ for brevity.  Its conformal dimension is given by
\begin{equation}
h \equiv h_{p'-1,1} = \frac{\brac{p'-2} \brac{p-2}}{4}
\end{equation}
and its fusion rules take the simple form
\begin{equation} \label{eqnSCFusion}
\phi \times \phi_{r,s} = \phi_{p'-r,s}.
\end{equation}
Note that $\phi \times \phi = \phi_{1,1}$.

We wish to extend the symmetry algebra of the minimal model $\MinMod{p'}{p}$ by adjoining the modes $\phi_n$ of the simple current to the Virasoro modes $L_m$.  Because the conformal dimension $h$ of the simple current is not integral in general, we do not expect that this extended algebra will be a Lie algebra, but will instead be defined by \gcrs{}.  Mathematically, we therefore seek a (graded) associative algebra $\alg{A}_{p',p}$ generated by the $\phi_n$, in which $\alg{Vir}$, or rather its \uea{} $\uealg{\alg{Vir}_{p',p}}$ appears as a (graded) subalgebra.  Here, the subscript ``$p',p$'' indicates that we identify $C$ with $c \id$ in this \uea{} (although we will often drop this subscript for brevity in what follows).

Before turning to the construction of this algebra, let us note that a part of the structure is already available to us.  As $\func{\phi}{w}$ is a primary field, its \ope{} with the energy-momentum tensor is
\begin{equation} \label{eqnOPEPrimary}
\radord{\func{T}{z} \func{\phi}{w}} = \frac{h \func{\phi}{w}}{\brac{z-w}^2} + \frac{\partial \func{\phi}{w}}{z-w} + \ldots
\end{equation}
Assuming that these two fields are mutually bosonic, this implies the familiar commutation rule
\begin{equation} \label{eqnCommLPhi}
\comm{L_m}{\phi_n} = \bigl( m \brac{h-1} - n \bigr) \phi_{m+n}.
\end{equation}
This (mutually) bosonic behaviour can in fact be derived, for example from the analogue of the Jacobi identity (see \cite[Eq.\ 2.5]{RidSU206}).

We can also extend the canonical antilinear antiautomorphism $L_m \rightarrow L_{-m}$ (defining the adjoint on representations) to the simple current modes.  The grading by conformal dimension restricts this extended adjoint to have the form $\phi_n \rightarrow \eps \phi_{-n}$, where $\abs{\eps} = 1$.  Requiring this to be an antiautomorphism with respect to \eqnref{eqnCommLPhi} gives no further constraints on $\eps$, so we may choose $\eps = 1$ for simplicity.  In other words, we choose the adjoint of the extended theory to be given by
\begin{equation} \label{eqnDefAdjoint}
L_m^{\dag} = L_{-m} \qquad \text{and} \qquad \phi_n^{\dag} = \phi_{-n}.
\end{equation}
We will verify later (\secref{secAlgebra}) that this defines an antilinear automorphism of the \emph{full} extended algebra $\alg{A}_{p',p}$, defined by \gcrs{}.

\section{Constructing the Extended Algebra} \label{secStructure}

\subsection{Commutativity and Associativity} \label{secAssoc}

We will suppose from now on that $p' \geqslant 3$, so that there always exists a unique non-trivial simple current $\phi$ of order $2$ and conformal dimension $h = \tfrac{1}{4} \brac{p'-2} \brac{p-2}$.  Since $\phi \times \phi = \phi_{1,1}$, the corresponding operator product expansion must be of the form
\begin{equation} \label{eqnNaiveOPE}
\radord{\func{\phi}{z} \func{\phi}{w}} = \sum_{j = 0}^{\infty} \func{A^{\brac{j}}}{w} \brac{z - w}^{j - 2h},
\end{equation}
where $\func{A^{\brac{0}}}{w}$ should be the identity field.  Observe that the exponent $2h$ need not be integral, signifying that we should take some care when using \eqnref{eqnNaiveOPE} in formal manipulations.

One of the requirements of the fields of a \cft{} is that they should satisfy some sort of \emph{mutual locality} principle, meaning that the \ope{} of two fields should be independent (up to a statistical phase) of the order in which the fields appear\footnote{We have already applied this principle in deriving \eqnref{eqnCommLPhi} from \eqnref{eqnOPEPrimary} --- there the fields were mutually \emph{bosonic}, meaning that the statistical phase factor is unity.}.  Applying this principle to the \ope{} (\ref{eqnNaiveOPE}) however leads to a slight ambiguity when $2h \notin \ZZ$:  Comparing the expansions of $\radord{\func{\phi}{z} \func{\phi}{w}}$ and $\radord{\func{\phi}{w} \func{\phi}{z}}$ is not possible unless we first agree on how to compare $\brac{z-w}^{j-2h}$ with $\brac{w-z}^{j-2h}$.

This issue may be bypassed by reformulating the mutual locality principle as the following commutativity requirement:
\begin{equation} \label{eqnCommutativity}
\brac{z - w}^{2h} \radord{\func{\phi}{z} \func{\phi}{w}} = \radord{\func{\phi}{w} \func{\phi}{z}} \brac{w - z}^{2h}.
\end{equation}
Inserting the expansion (\ref{eqnNaiveOPE}) and its $z \leftrightarrow w$ analogue into this equation now leads to integral powers of $z-w$ and $w-z$, settling the ambiguity mentioned above.  We mention that in considering the mutual locality of two different fields (such as the construction in \cite{RidSU206}), we need only require that the corresponding equality in \eqnref{eqnCommutativity} hold up to a phase factor.  However, it is easy to check that when the two fields are identical (the case of interest here), this phase factor is necessarily unity.

Another requirement of the \ope{} of fields is that it defines an associative operation.  Assuming this requirement, the triple product $\radord{\func{\phi}{x} \func{\phi}{z} \func{\phi}{w}}$ is unambiguously defined.  We now derive an interesting conclusion by combining this requirement with the commutativity condition, \eqnref{eqnCommutativity}, and the generic \ope{} (\ref{eqnNaiveOPE}).

Applying commutativity twice, we may write
\begin{equation}
\brac{x-w}^{2h} \brac{z-w}^{2h} \radord{\func{\phi}{x} \func{\phi}{z} \func{\phi}{w}} = \radord{\func{\phi}{w} \func{\phi}{x} \func{\phi}{z}} \brac{w-x}^{2h} \brac{w-z}^{2h}.
\end{equation}
Expanding $\radord{\func{\phi}{x} \func{\phi}{z}}$ as in \eqnref{eqnNaiveOPE}, multiplying both sides by $\brac{x-z}^{2h - \gamma}$ (for some arbitrary $\gamma \in \ZZ$), and contour-integrating $x$ around $z$ gives
\begin{equation}
\sum_{j=0}^{\gamma - 1} \binom{2h}{\gamma - j - 1} \sqbrac{\func{A^{\brac{j}}}{z} \func{\phi}{w} \brac{z-w}^{4h + j - \gamma + 1} - \brac{-1}^{j - \gamma + 1} \func{\phi}{w} \func{A^{\brac{j}}}{z} \brac{w-z}^{4h + j - \gamma + 1}} = 0.
\end{equation}
As $4h \in \ZZ$, this may be simplified to
\begin{equation}
\sum_{j=0}^{\gamma - 1} \binom{2h}{\gamma - j - 1} \brac{z-w}^{4h + j - \gamma + 1} \sqbrac{\func{A^{\brac{j}}}{z} \func{\phi}{w} - \brac{-1}^{4h} \func{\phi}{w} \func{A^{\brac{j}}}{z}} = 0,
\end{equation}
from which we may conclude that
\begin{equation} \label{eqnCommAPhi}
\func{A^{\brac{j}}}{z} \func{\phi}{w} = \brac{-1}^{4h} \func{\phi}{w} \func{A^{\brac{j}}}{z},
\end{equation}
for each $j$, by analysing $\gamma = 1 , 2 , 3 , \ldots$ consecutively.  This proves that the fields $\func{A^{\brac{j}}}{z}$ appearing in the \ope{} (\ref{eqnNaiveOPE}) commute with $\func{\phi}{w}$ when $2h \in \ZZ$ (the fields are mutually bosonic), but anticommute with $\func{\phi}{w}$ when $2h \notin \ZZ$ (the fields are mutually fermionic).

Recall now that $\func{A^{\brac{0}}}{z}$ was supposed to be the identity field.  This clearly contradicts the above commutativity conclusion when $2h \notin \ZZ$.  Indeed, the $\func{A^{\brac{j}}}{z}$ with $j > 0$ should be (Virasoro) descendants of the identity field, hence should be expressible in terms of normally-ordered products of $\func{T}{z}$ and its derivatives.  But these are also mutually bosonic with respect to $\func{\phi}{w}$ (\secref{secNotation}), so we face a similar contradiction.  This contradiction can only be satisfactorily resolved if we assume that each $\func{A^{\brac{j}}}{z}$ contains an operator $\mathcal{S}$ which satisfies
\begin{equation} \label{eqnSComm}
\mathcal{S} \func{T}{z} = \func{T}{z} \mathcal{S}, \qquad \text{but} \qquad \mathcal{S} \func{\phi}{w} = \brac{-1}^{4h} \func{\phi}{w} \mathcal{S}.
\end{equation}
Such an operator $\mathcal{S}$ was first introduced in \cite{JacQua06} for the $\MinMod{3}{p}$ models (though with a less direct justification).

It is worth noting that associativity does \emph{not} require the introduction of $\mathcal{S}$-type operators in the corresponding full non-chiral theories (with the diagonal modular invariant) \cite{DotOpe85}.  Essentially, the antiholomorphic component will contribute an additional factor of $\brac{-1}^{4h}$ to \eqnref{eqnCommAPhi}, removing the contradiction that necessitated the appearance of $\mathcal{S}$.  Whilst it may then be argued that this operator is in some sense ``unphysical'', the fact remains that the discipline of \cft{} rests heavily upon its chiral foundations, and many of its applications (especially in mathematics, but also physically when boundaries are concerned) require a consistent chiral formulation.  We will illustrate this explicitly in \secref{secExamples}, by demonstrating that the properties of $\mathcal{S}$ are \emph{crucial} in showing that the $\MinMod{3}{5}$ singular vectors vanish identically.

We will find it convenient to explicitly factor this operator $\mathcal{S}$ out from each of the $\func{A^{\brac{j}}}{z}$, redefining the latter so that the operator product expansion is 
\begin{equation} \label{eqnOPEGeneral}
\radord{\func{\phi}{z} \func{\phi}{w}} = \mathcal{S} \sum_{j = 0}^{\infty} \func{A^{\brac{j}}}{w} \brac{z - w}^{j - 2h}.
\end{equation}
This expansion replaces \eqnref{eqnNaiveOPE}, to which we shall not refer again.  Thus, for example, $\func{A^{\brac{0}}}{w}$ is now genuinely the identity field.  We note that $\mathcal{S}^2$ commutes with $\func{\phi}{z}$, hence with its modes, and is therefore a multiple of the identity in any irreducible module of the extended algebra (which will be constructed shortly).  We also note that $\mathcal{S}$ must leave the vacuum $\ket{0}$ invariant, so that it does not interfere with the state-field correspondence.

\subsection{Algebraic Structure} \label{secAlgebra}

We turn now to the derivation of the algebra defined by the modes of $\func{\phi}{z}$.  At a formal level, this uses a standard trick \cite{ZamNon85} involving the evaluation of
\begin{equation} \label{eqnRDefined}
\func{R_{m,n}}{\gamma} = \oint_0 \oint_w \radord{\func{\phi}{z} \func{\phi}{w}} z^{m - h + \gamma - 1} w^{n + h - 1} \brac{z - w}^{2h - \gamma} \frac{\dd z}{2 \pi \ii} \frac{\dd w}{2 \pi \ii} \qquad \text{($\gamma \in \ZZ$)}
\end{equation}
in two distinct ways.  We can expand the operator product directly, using \eqnref{eqnOPEGeneral}, or we can break the $z$-contour around $w$ into the difference of two contours about the origin, one with $\abs{z} > \abs{w}$ and the other with $\abs{z} < \abs{w}$.  The final result is a \gcr{}, parametrised by $\gamma$, $m$ and $n$:
\begin{equation} \label{eqnGCRGeneral}
\sum_{\ell = 0}^{\infty} \binom{\ell - 2h + \gamma - 1}{\ell} \Big[ \phi_{m-\ell} \phi_{n+\ell} - \brac{-1}^{\gamma} \phi_{n + 2h - \gamma - \ell} \phi_{m - 2h + \gamma + \ell} \Big] = \mathcal{S} \sum_{j=0}^{\gamma-1} \binom{m - h + \gamma - 1}{\gamma - 1 - j} A_{m+n}^{\brac{j}},
\end{equation}
where $A_{m+n}^{\brac{j}}$ denotes the modes of the fields $\func{A^{\brac{j}}}{w}$ appearing in the \ope{} (\ref{eqnOPEGeneral}).  Note that $\gamma$ determines how many terms of this \ope{} contribute to the corresponding \gcr{}.

We observe that if $2h \in \ZZ$, the \gcrs{} with $\gamma = 2h$ reduce to commutation or anticommutation relations:
\begin{equation}
\phi_m \phi_n - \brac{-1}^{2 h} \phi_n \phi_m = \mathcal{S} \sum_{j=0}^{2h-1} \binom{m+h-1}{2h-1-j} A_{m+n}^{\brac{j}}.
\end{equation}
In contrast, when $2h \notin \ZZ$, every \gcr{} has infinitely many terms on the left hand side.

It remains to determine the fields $\func{A^{\brac{j}}}{w}$, at least for small $j$.  As in \cite{RidSU206}, we invert \eqnref{eqnOPEGeneral},
\begin{equation}
\mathcal{S} \func{A^{\brac{j}}}{w} = \oint_w \radord{\func{\phi}{z} \func{\phi}{w}} \brac{z - w}^{2h - j - 1} \frac{\dd z}{2 \pi \ii},
\end{equation}
let both sides act on the vacuum, and send $w$ to $0$.  The result determines the corresponding states as
\begin{equation} \label{eqnDefAStates}
\ket{A^{\brac{j}}} = \mathcal{S} \ket{A^{\brac{j}}} = \phi_{h-j} \phi_{-h} \ket{0} = \phi_{h-j} \ket{\phi}.
\end{equation}
[The fields $\func{A^{\brac{j}}}{w}$ may be expressed purely in terms of $\func{T}{w}$ and its derivatives, hence $\ket{A^{\brac{j}}}$ is a linear combination of the grade $j$ Virasoro descendants of the vacuum, and so $\mathcal{S} \ket{A^{\brac{j}}} = \ket{A^{\brac{j}}}$.]

There is no such descendant at grade $1$, so $\ket{A^{\brac{1}}}$ = 0, hence $\func{A^{\brac{1}}}{w} = 0$.  There is a unique Virasoro descendant at grade $2$, so $\ket{A^{\brac{2}}}$ must be proportional to $L_{-2} \ket{0}$.  This constant of proportionality may be evaluated by using \eqnDref{eqnCommLPhi}{eqnVirComm} to derive
\begin{equation}
\bra{0} L_2 \ket{A^{\brac{2}}} = \bra{0} L_2 \phi_{h-2} \phi_{-h} \ket{0} = h \qquad \text{and} \qquad \bra{0} L_2 L_{-2} \ket{0} = \frac{c}{2}.
\end{equation}
It follows that that $\func{A^{\brac{2}}}{w} = \tfrac{2h}{c} \func{T}{w}$.  Similarly, $\func{A^{\brac{3}}}{w} = \frac{h}{c} \partial \func{T}{w}$.

At grade $4$, there are two possible Virasoro descendants, so we have $\ket{A^{\brac{j}}} = \alpha L_{-4} \ket{0} + \beta L_{-2}^2 \ket{0}$, for some unknowns $\alpha$ and $\beta$.  Applying $\bra{0} L_4$ and $\bra{0} L_2^2$ to both sides gives two linear equations to solve, which may be solved to give
\begin{equation}
\func{A^{\brac{4}}}{w} = \frac{3 h \brac{c - 2h + 4}}{2 c \brac{5c + 22}} \partial^2 \func{T}{w} + \frac{2 h \brac{5h + 1}}{c \brac{5c + 22}} \normord{\func{T}{w} \func{T}{w}}.
\end{equation}
Of course these expressions are all well-known \cite{ZamInf85}.

We remark that when $c = \frac{-22}{5}$, this last computation breaks down because the vacuum $\alg{Vir}$-module has a singular vector at grade $4$.  As far as the inner-product is concerned, $L_{-4} \ket{0}$ and $L_{-2}^2 \ket{0}$ are not linearly independent.  We would therefore need to compute $\ket{A^{\brac{4}}}$ separately for this special value of the central charge (choosing which state we regard as independent).  Of course, this central charge corresponds to the $\MinMod{2}{5}$ model, which is not of interest here, as it has no non-trivial simple current.  However, singular vectors will eventually appear (for example, at grade $6$ for $\MinMod{3}{4}$) and therefore complicate the calculations.

It should be clear then that a knowledge of the singular vectors of the Virasoro vacuum module is essential to continuing these derivations.  Whilst this knowledge is not beyond reach \cite{BenDeg88}, the implementation rapidly becomes cumbersome at higher grades.  It is therefore worth observing that it is possible to derive recurrence relations for the $\func{A^{\brac{j}}}{w}$.  We will present such derivations as an application of our extended algebra formalism in \secref{secOPERecursive}.

For later reference, we display a few of the \gcrs{}.  When applying these relations, it is often useful to explicitly indicate the value of $\gamma$ employed.  As in \cite{RidSU206}, we will use ``$\gcreq{\gamma}$'' to indicate an equality obtained using a \gcr{} with parameter $\gamma$.  Clearly when $\gamma \leqslant 0$, the right-hand-side of the \gcrs{} vanish, giving
\begin{equation}
\sum_{\ell = 0}^{\infty} \binom{\ell - 2h + \gamma - 1}{\ell} \Big[ \phi_{m-\ell} \phi_{n+\ell} - \brac{-1}^{\gamma} \phi_{n + 2h - \gamma - \ell} \phi_{m - 2h + \gamma + \ell} \Big] \gcreq{\gamma \leqslant 0} 0.
\end{equation}
Similarly, $\func{A^{\brac{0}}}{w} = 1$ implies that the $\gamma = 1$ generalised conjugation relation is
\begin{equation}
\sum_{\ell = 0}^{\infty} \binom{\ell - 2h}{\ell} \Big[ \phi_{m-\ell} \phi_{n+\ell} + \phi_{n + 2h - 1 - \ell} \phi_{m - 2h + 1 + \ell} \Big] \gcreq{1} \delta_{m+n,0} \mathcal{S}.
\end{equation}
Continuing, we have:
\begin{equation}
\sum_{\ell = 0}^{\infty} \binom{\ell - 2h + 1}{\ell} \Big[ \phi_{m-\ell} \phi_{n+\ell} - \phi_{n + 2h - 2 - \ell} \phi_{m - 2h + 2 + \ell} \Big] \gcreq{2} -\brac{n+h-1} \delta_{m+n,0} \mathcal{S},
\end{equation}
\begin{equation}
\sum_{\ell = 0}^{\infty} \binom{\ell - 2h + 2}{\ell} \Big[ \phi_{m-\ell} \phi_{n+\ell} + \phi_{n + 2h - 3 - \ell} \phi_{m - 2h + 3 + \ell} \Big] \gcreq{3} \sqbrac{\binom{n+h-1}{2} \delta_{m+n,0} + \frac{2h}{c} L_{m+n}} \mathcal{S},
\end{equation}
\begin{multline}
\sum_{\ell = 0}^{\infty} \binom{\ell - 2h + 3}{\ell} \Big[ \phi_{m-\ell} \phi_{n+\ell} - \phi_{n + 2h - 4 - \ell} \phi_{m - 2h + 4 + \ell} \Big] \\*
\gcreq{4} \sqbrac{-\binom{n+h-1}{3} \delta_{m+n,0} + \frac{h}{c} \brac{m - n - 2h + 4} L_{m+n}} \mathcal{S},
\end{multline}
\begin{multline}
\text{and} \quad \sum_{\ell = 0}^{\infty} \binom{\ell - 2h + 4}{\ell} \Big[ \phi_{m-\ell} \phi_{n+\ell} + \phi_{n + 2h - 5 - \ell} \phi_{m - 2h + 5 + \ell} \Big] \gcreq{5} \Biggl[ \binom{n+h-1}{4} \delta_{m+n,0} \Biggr. \\*
+ \sqbrac{\frac{3 h \brac{c - 2h + 4}}{c \brac{5 c + 22}} \binom{m+n+3}{2} - \frac{h}{c} \brac{m-h+4} \brac{n+h-1}} L_{m+n} \\*
\Biggl. + \frac{2 h \brac{5 h + 1}}{c \brac{5 c + 22}} \sum_{r \in \ZZ} \normord{L_r L_{m+n-r}} \Biggr] \mathcal{S}.
\end{multline}

We now define\footnote{This definition is in fact not quite complete because we have thus far avoided specifying the values that the index in $\phi_n$ can take.  As with all fields exhibiting non-bosonic statistics, these values depend upon the state on which the mode acts.  This will be specified when discussing the representation theory of the symmetry algebra in \secref{secRepTheory}.} the extended symmetry algebra $\alg{A}_{p',p}$ of the minimal model $\MinMod{p'}{p}$ (with $p' > 2$) to be the graded (by conformal dimension) associative algebra generated by the modes $\phi_n$, subject to the set of \gcrs{}, \eqnref{eqnGCRGeneral}, and equipped with the adjoint $\phi_n^{\dag} = \phi_{-n}$.  It is easy to check that $\func{R_{m,n}}{\gamma}^{\dag} = \func{R_{-n,-m}}{\gamma}$ (using the left-hand-side of \eqnref{eqnGCRGeneral}), hence that this adjoint defines a genuine antilinear antiautomorphism of $\alg{A}_{p',p}$.  We summarise this result by noting that this adjoint makes our extended symmetry algebra into a graded $^*$-algebra (the grading being by the conformal weight and the $^*$-algebra meaning simply that the adjoint satisfies the usual properties with respect to the algebra operations).  Note that the \gcr{} with $\gamma = 1$ requires $\mathcal{S}$ to be self-adjoint.

\section{Representation Theory} \label{secRepTheory}

\subsection{Monodromy Charge} \label{secMonCharge}

Consider now the Virasoro highest weight state $\ket{\phi_{r,s}}$ corresponding to the primary field $\func{\phi_{r,s}}{z}$.  The simple current fusion rules (\ref{eqnSCFusion}) imply that
\begin{equation} \label{eqnOPEPhiOther}
\radord{\func{\phi}{z} \func{\phi_{r,s}}{w}} = \frac{\eta_{r,s} \func{\phi_{p'-r,s}}{w}}{\brac{z-w}^{\theta_{r,s}}} + \ldots,
\end{equation}
where $\eta_{r,s}$ is a constant (or possibly an operator like $\mathcal{S}$ if required by associativity).  The leading exponent $\theta_{r,s}$ is then given by
\begin{equation} \label{eqnDefCharge}
\theta_{r,s} = h + h_{r,s} - h_{p'-r,s} = 1 - rs + \frac{1}{2} \sqbrac{p \brac{r-1} + p' \brac{s-1}} \in \frac{1}{2} \ZZ.
\end{equation}
As $\phi$ is a simple current, the omitted terms in the \ope{} (\ref{eqnOPEPhiOther}) all have $\brac{z-w}$-exponents of the form $j - \theta_{r,s}$, where $j \in \ZZ_+$.

The common value (modulo $1$) of these exponents tells us how to expand $\func{\phi}{z}$ into modes, when acting on $\ket{\phi_{r,s}}$.  We see this by noting that
\begin{equation}
\func{\phi}{z} \ket{\phi_{r,s}} = \lim_{w \rightarrow 0} \radord{\func{\phi}{z} \func{\phi_{r,s}}{w}} \ket{0} = \eta_{r,s} z^{-\theta_{r,s}} \ket{\phi_{p'-r,s}} + \ldots
\end{equation}
implies that $\func{\phi}{z}$ must be expanded in powers of $z$ equal to $-\theta_{r,s}$ modulo $1$:
\begin{equation} \label{eqnModeDecomposition}
\func{\phi}{z} \ket{\phi_{r,s}} = \sum_{n \in \ZZ + \theta_{r,s} - h} \phi_n z^{-n - h} \ket{\phi_{r,s}}.
\end{equation}
We interpret this as investing each highest weight state with an associated charge, $\theta_{r,s}$, whose value modulo $1$ dictates which modes of the simple current field may act upon this state.  Modes corresponding to the wrong charge do not have a well-defined action on this state.  We shall refer to $\theta_{r,s}$ as the $\func{\alg{u}}{1}$-charge of $\ket{\phi_{r,s}}$, and we shall call its value modulo $1$ the \emph{monodromy charge} (following \cite{SchSim90}).

Whilst the monodromy charge controls which indices on the modes $\phi_n$ are allowed when acting on a \hws{}, the $\func{\alg{u}}{1}$-charge tells us for which indices this action necessarily gives $0$.  By writing $\phi_n \ket{\phi_{r,s}}$ as a contour integral involving the \ope{} (\ref{eqnOPEPhiOther}), it is easy to see when the integrand becomes regular (hence when the integral vanishes).  The result is
\begin{equation} \label{eqnFirstDescendant}
\phi_n \ket{\phi_{r,s}} = 0 \qquad \text{for all $n > \theta_{r,s} - h$,}
\end{equation}
and of course this is non-vanishing when $n = \theta_{r,s} - h$.  In other words, the $\func{\alg{u}}{1}$-charge specifies the \emph{first descendant} of the state $\ket{\phi_{r,s}}$ with respect to the extended algebra $\alg{A}_{p',p}$.

The $\func{\alg{u}}{1}$-charge is non-negative on the Virasoro highest weight states $\ket{\phi_{r,s}}$.  To prove this, we note that as $\gcd \set{p' , p} = 1$, either $p$ or $p'$ must be odd.  Without loss of generality, we suppose it is $p$ (otherwise swap $p$ and $p'$, and $r$ and $s$ in what follows).  Then, for each $s = 1 , \ldots , p-1$, there is a unique $r \in \RR$ such that $\theta_{r,s} = 0$.  From \eqnref{eqnDefCharge}, this value is
\begin{equation}
r = 1 + \frac{\brac{p'-2} \brac{s-1}}{2s-p}.
\end{equation}
Suppose that this unique $r$ lies between $1$ and $p' - 1$ (non-inclusive).  Then
\begin{equation}
0 < \frac{s-1}{2s-p} < 1, \qquad \text{as $p' \geqslant 3$.}
\end{equation}
If $s > p/2$, this requires $s > p-1$.  Similarly, if $s < p/2$, this requires $s < 1$, so both conclusions fall outside the allowed range for $s$.  It therefore follows that for $s = 1 , \ldots , p-1$, $\theta_{r,s} \neq 0$ for $1 < r < p' - 1$.  Now, $\theta_{1,s} = \theta_{p'-1 , p-s} = \frac{1}{2} \brac{p'-2} \brac{s-1} > 0$ for all $s \neq 1$, so we conclude that $\theta_{r,s} \geqslant 0$ with equality if and only if $\phi_{r,s}$ is the identity.
 
We can generalise the notion of monodromy charge to descendant states in the same fashion.  It is not difficult to infer (as in \cite{RidSU206}) from \eqnDref{eqnCommLPhi}{eqnGCRGeneral} that the monodromy charge of a state is left invariant (modulo $1$) by the application of a Virasoro mode $L_m$, but is changed by $2h$ (again modulo $1$) by the application of a mode $\phi_n$.

\subsection{$\alg{A}_{p',p}$-Verma Modules} \label{secVerma}

We define an $\alg{A}_{p',p}$-\hws{} to be a state $\ket{\psi}$ satisfying
\begin{equation} \label{eqnDefHWS}
\phi_n \ket{\psi} = L_m \ket{\psi} = 0 \qquad \text{for all $m,n > 0$}.
\end{equation}
We include annihilation under the Virasoro modes of positive index to ensure that an $\alg{A}_{p',p}$-\hws{} is necessarily\footnote{The Virasoro highest weight condition follows easily from $\phi_n \ket{\psi} = 0$ when $h$ is sufficiently small (use the \gcr{} with $\gamma = 4$ for example).  However, it is not clear if this continues to hold true for all $h$.} a Virasoro-\hws{}.  An $\alg{A}_{p',p}$-Verma module is then the module generated from such a highest weight state by the action of the (allowed) $\phi_n$, modulo the algebra relations (the \gcrs{}).  We will denote the $\alg{A}_{p',p}$-Verma module generated from the \hws{} $\ket{\phi_{r,s}}$ by $\ExtVerMod{p',p}{r,s}$.

Consider therefore an arbitrary $\alg{A}_{p',p}$-\hws{} $\ket{\psi}$.  Being a $\alg{Vir}$-\hws{}, it has (\secref{secMonCharge}) a definite $\func{\alg{u}}{1}$-charge $\theta$.  Its first descendant is then given by \eqnref{eqnFirstDescendant} as
\begin{equation}
\phi_{\theta - h} \ket{\psi} \neq 0.
\end{equation}
If $\theta > h$, this contradicts the definition of \hws{} given in \eqnref{eqnDefHWS}, so we conclude that $\alg{A}_{p',p}$-\hwss{} must necessarily have $\func{\alg{u}}{1}$-charge $\theta \leqslant h$.

Now recall that the Virasoro \hwss{} $\ket{\phi_{r,s}}$ appearing in the minimal models have $\theta_{r,s} \geqslant 0$ and $\theta_{r,s} \in \frac{1}{2} \ZZ$.  It is easy to derive the identity
\begin{equation}
\theta_{r,s} + \theta_{p'-r,s} = \frac{1}{2} \brac{p'-2} \brac{p-2} = 2h
\end{equation}
from \eqnref{eqnDefCharge}, from which we deduce that $\theta_{r,s} \leqslant h$ if and only if $\theta_{p'-r,s} \geqslant h$.  In other words, $\ket{\phi_{r,s}}$ is an $\alg{A}_{p',p}$-\hws{} precisely when $\ket{\phi_{p'-r,s}}$ is not (unless $\theta_{r,s} = h$, a special case that we shall discuss shortly).

It follows that these $\alg{Vir}$-\hwss{} of charge greater than $h$ must occur as descendants (with respect to the extended algebra) of the $\alg{A}_{p',p}$-\hwss{}, whose charges are not more than $h$.  We specify this relationship precisely by considering the (first) descendant state $\phi_{\theta_{r,s} - h} \ket{\phi_{r,s}}$ (where $\theta_{r,s} \leqslant h$).  The conformal dimension of this descendant is $h_{r,s} - \theta_{r,s} + h = h_{p'-r,s}$, by \eqnref{eqnDefCharge}, suggesting that
\begin{equation} \label{eqnIdentVirHWS}
\phi_{\theta_{r,s} - h} \ket{\phi_{r,s}} = \ket{\phi_{p'-r,s}},
\end{equation}
the $\alg{Vir}$-\hws{}.  This can be confirmed by acting with the positive Virasoro modes ($n > 0$) and using \eqnref{eqnFirstDescendant}:
\begin{equation}
L_n \phi_{\theta_{r,s} - h} \ket{\phi_{r,s}} = \comm{L_n}{\phi_{\theta_{r,s} - h}} \ket{\phi_{r,s}} = \brac{n \brac{h-1} - \theta_{r,s} + h} \phi_{n + \theta_{r,s} - h} \ket{\phi_{r,s}} = 0.
\end{equation}

We remark that it is possible for the \hws{} $\ket{\phi_{r,s}}$ to have $\theta_{r,s} = h$.  In fact, this occurs if and only if $r = p' / 2$ or $s = p / 2$ (hence cannot occur if $p$ and $p'$ are both odd).  We have then two $\alg{Vir}$-\hwss{}, $\ket{\phi_{r,s}}$ and $\phi_0 \ket{\phi_{r,s}}$, of the same conformal dimension.  These are both $\alg{A}_{p',p}$-\hwss{}, according to the above definition.  We now ask whether these two \hwss{} are linearly independent.  A relevant observation to this question is that as $\phi_0 \ket{\phi_{r,s}}$ is a $\alg{Vir}$-\hws{},
\begin{equation} \label{eqnPhi0Eig}
\bracket{\phi_{r,s}}{\phi_0^2}{\phi_{r,s}} = 1, \qquad \text{so} \qquad \phi_0^2 \ket{\phi_{r,s}} = \ket{\phi_{r,s}},
\end{equation}
assuming $\ket{\phi_{r,s}}$ and $\phi_0 \ket{\phi_{r,s}}$ to be either proportional or orthogonal.

If we take these two \hwss{} to be proportional, then \eqnref{eqnPhi0Eig} limits the proportionality constant to $\pm 1$.  We therefore have two possible $\alg{A}_{p',p}$-Verma modules which are identical as $\alg{Vir}$-Verma modules, but which are distinguished by the eigenvalue of $\phi_0$ on their \hwss{}.  If we instead take these two \hwss{} to be orthogonal, then the linear combinations
\begin{equation} \label{eqnPhi0EigVects}
\ket{\phi_{r,s}} \pm \phi_0 \ket{\phi_{r,s}}
\end{equation}
are the \hwss{} of (distinct) $\alg{A}_{p',p}$-Verma modules.  Again, these are identical as $\alg{Vir}$-Verma modules, but can distinguished by the eigenvalue of $\phi_0$ on their \hwss{}.  It follows that the choice of whether $\phi_0 \ket{\phi_{r,s}}$ is proportional or orthogonal to $\ket{\phi_{r,s}}$ is of no essential importance.

The above discussion suggests that we should augment the definition of an $\alg{A}_{p',p}$-\hws{} to include being an eigenstate of $\phi_0$ (assuming that $\phi_0$ is allowed to act upon it).  The eigenvalue, when defined, would then be $\pm 1$ for states of $\func{\alg{u}}{1}$-charge $h$, and $0$ otherwise.  Whilst this is in full accord with general Lie-algebraic principles\footnote{Here we mean that $\phi_0$ commutes with $L_0$, $C$ and $\mathcal{S}$ whenever it belongs to $\alg{A}_{p',p}$, hence may be consistently included in a ``Cartan subalgebra'' (maximal abelian subalgebra) of $\alg{A}_{p',p}$.}, considering eigenstates of $\phi_0$ leads to a certain inelegance in the formalism.  In particular, we are forced to relinquish our picture of the $\phi_n$ as being \emph{intertwiners} between $\alg{Vir}$-Verma modules, that is that the action of each $\phi_m$ takes us from one $\alg{Vir}$-Verma module to another whilst the action of a subsequent $\phi_n$ brings us back again.

To restore this intertwining picture, we shall adopt the following convention concerning modules whose \hwss{} have $\func{\alg{u}}{1}$-charge $h$:  We declare that $\ket{\phi_{r,s}}$ is an $\alg{A}_{p',p}$-\hws{} and that $\phi_0 \ket{\phi_{r,s}}$ is its orthogonal descendant.  In this way, $\phi_0$ (as well as the other $\phi_n$) act as genuine intertwiners between the $\alg{Vir}$-Verma modules generated from $\ket{\phi_{r,s}}$ and $\phi_0 \ket{\phi_{r,s}}$.  The $\alg{A}_{p',p}$-Verma module\footnote{This module however has the curious property of decomposing into \emph{two} different $\alg{A}_{p',p}$-Verma modules, headed by the vectors (\ref{eqnPhi0EigVects}).  This is indicative of the fact that this module is not, strictly speaking, a Verma module, because its \hws{} is not an eigenstate of the \emph{maximal} abelian subalgebra of $\alg{A}_{p',p}$:  We have deliberately chosen a \hws{} which is not an eigenvector of $\phi_0$.  Nevertheless, we will refer to this module as a Verma module (for regularity in exposition), with this slight subtlety understood implicitly.} generated by $\ket{\phi_{r,s}}$ then contains two $\alg{Vir}$-\hwss{} (given above), just as the $\alg{A}_{p',p}$-Verma modules with $\func{\alg{u}}{1}$-charge not equal to $h$ do.

The picture which now emerges is that $\alg{A}_{p',p}$-Verma modules have a \hws $\ket{\phi_{r,s}}$ which is a $\alg{Vir}$-\hws{} with $\theta_{r,s} \leqslant h$, and the first descendant $\phi_{\theta_{r,s} - h} \ket{\phi_{r,s}}$ is another $\alg{Vir}$-\hws{} $\ket{\phi_{p'-r,s}}$ which has $\theta_{p'-r,s} \geqslant h$.  In fact, this exhausts the set of (independent) Virasoro highest weight states (which are not themselves Virasoro descendants) in any $\alg{A}_{p',p}$-Verma module.  The proof of this statement is identical to that of the analogous statement in \cite[Prop.\ 5.2]{RidSU206}, so we omit it here.

An $\alg{A}_{p',p}$-Verma module therefore decomposes into two $\alg{Vir}$-modules.  From a mathematical perspective, what we have shown is that the \emph{injection} of graded $^*$-algebras,
\begin{equation}
\uealg{\alg{Vir}_{p',p}} \longrightarrow \alg{A}_{p',p},
\end{equation}
describing the extension of the symmetry algebra, leads to a \emph{surjection} of the corresponding Verma modules:
\begin{equation} \label{Surjection}
\VirVerMod{p',p}{r,s} \oplus \VirVerMod{p',p}{p'-r,s} \longrightarrow \ExtVerMod{p',p}{r,s}.
\end{equation}
Here, $\VirVerMod{p',p}{r,s}$ denotes the $\alg{Vir}_{p',p}$-Verma module whose \hws{} is $\ket{\phi_{r,s}}$.  We mention that this surjection is a homomorphism of $\alg{Vir}$-modules (strictly speaking, $\alg{Vir} \oplus \alg{Vir}$-modules), meaning nothing more than that the action of the Virasoro modes is the same on both sides.  Furthermore, because the adjoints of these $^*$-algebras completely determine the sesquilinear forms on the Verma modules up to normalisation, it follows that we can choose (and indeed have chosen) this normalisation so that our surjection is \emph{isometric} (norm-preserving).

\subsection{$\mathcal{S}$-eigenvalues} \label{secSEig}

Recall from \secref{secAssoc} that $\mathcal{S}$ commutes with all the $\phi_n$ up to a factor of $\brac{-1}^{4h}$.  Its action on an $\alg{A}_{p',p}$-Verma module is therefore completely determined by its eigenvalue on the highest weight state.  Using the generalised commutation relations, \eqnref{eqnGCRGeneral}, we will compute (some of) these eigenvalues, under the assumption that every $\alg{Vir}$-\hws{} has norm $1$.  We will describe an algorithm for computing \emph{all} these eigenvalues (recursively) in \secref{secSEigRecursive}.

On a $\alg{A}_{p',p}$-\hws{} of $\func{\alg{u}}{1}$-charge $0$ (which is necessarily the vacuum $\ket{0}$), the generalised commutation relation with $\gamma = 1$ gives
\begin{equation} \label{eqnSEigCh=0}
\bra{0} \mathcal{S} \ket{0} \gcreq{1} \bra{0} \phi_h \phi_{-h} \ket{0} = \braket{\phi}{\phi} = 1 \quad \Rightarrow \quad \mathcal{S} \ket{0} = \ket{0}.
\end{equation}
This is of course a consequence of the normalisation we assumed at the end of \secref{secAssoc}.  However, if $\ket{\psi}$ is a $\alg{A}_{p',p}$-\hws{} of charge $\frac{1}{2}$, then we derive
\begin{equation} \label{eqnSEigCh=1/2}
\bra{\psi} \mathcal{S} \ket{\psi} \gcreq{1} \bra{\psi} 2 \phi_{h - 1/2} \phi_{1/2 - h} \ket{\psi} = 2 \quad \Rightarrow \quad \mathcal{S} \ket{\psi} = 2 \ket{\psi},
\end{equation}
as $\phi_{1/2 - h} \ket{\psi}$ is a (normalised) Virasoro \hws{}.  Interestingly, charge $1$ $\alg{A}_{p',p}$-\hwss{} $\ket{\psi}$ require $\gamma = 3$:
\begin{equation} \label{eqnSEigCh=1}
1 = \bra{\psi} \phi_{h - 1} \phi_{1 - h} \ket{\psi} \gcreq{3} \bra{\psi} \frac{2h}{c} L_0 \mathcal{S} \ket{\psi} \quad \Rightarrow \quad \mathcal{S} \ket{\psi} = \frac{c}{2 h h_{\psi}} \ket{\psi},
\end{equation}
where $h_{\psi}$ is the conformal dimension of $\ket{\psi}$, as does the charge $\frac{3}{2}$ case:
\begin{equation} \label{eqnSEigCh=3/2}
1 = \bra{\psi} \phi_{h - 3/2} \phi_{3/2 - h} \ket{\psi} \gcreq{3} \bra{\psi} \frac{1}{2} \brac{\frac{-1}{8} + \frac{2h}{c} L_0} \mathcal{S} \ket{\psi} \quad \Rightarrow \quad \mathcal{S} \ket{\psi} = \frac{16 c}{16 h h_{\psi} - c} \ket{\psi}.
\end{equation}
Similarly, the charge $2$ case requires $\gamma = 5$:
\begin{gather}
1 = \bra{\psi} \phi_{h - 2} \phi_{2 - h} \ket{\psi} \gcreq{5} \bra{\psi} \sqbrac{ \frac{-h \brac{c + 18 h + 8}}{c \brac{5c + 22}} L_0 + \frac{2 h \brac{5h + 1}}{c \brac{5c + 22}} \brac{L_0^2 + 2 L_0}} \mathcal{S} \ket{\psi} \notag \\
\Rightarrow \quad \mathcal{S} \ket{\psi} = \frac{c \brac{5c + 22}}{h h_{\psi} \sqbrac{2 \brac{5h + 1} h_{\psi} - \brac{c - 2h + 4}}} \ket{\psi}, \label{eqnSEigCh=2}
\end{gather}
and so on.

We remark that we could have instead chosen the norms of the $\alg{Vir}$-\hwss{} (or included constant factors in \eqnref{eqnIdentVirHWS} appropriately) so as to make the $\mathcal{S}$-eigenvalues above equal to $1$.  However, we would still have to compute these norms in any genuine calculation, so this does not represent a simplification of the formalism.  We observe that to compute the $\mathcal{S}$-eigenvalue of a $\alg{A}_{p',p}$-\hws{} of charge $\theta$, it is necessary to use a $\gamma = 2 \theta + 1$ \gcr{} when $\theta \in \ZZ$, but we can get away with a $\gamma = 2 \theta$ relation when $\theta \notin \ZZ$.  This pattern regarding the required orders of \gcrs{} is quite common when computing with extended algebras, and we shall see why in \secref{secRSymm}.

Consider now a theory with $h \in \ZZ$, $\MinMod{3}{10}$ or $\MinMod{5}{6}$ for example.  According to the above discussion, computing the eigenvalue of $\mathcal{S}$ on an $\alg{A}_{p',p}$-module of charge $\theta = h$ requires us to employ a generalised commutation relation with $\gamma = 2h + 1$ (and no smaller).  This is interesting as this relation receives contributions from every singular term of the operator product expansion (\ref{eqnOPEGeneral}), as well as from the first regular term.  In other words, it is not possible to compute this $\mathcal{S}$-eigenvalue using only the information contained in the singular terms.  We will shortly see further computations where it will prove necessary to use \gcrs{} which include contributions from further regular terms of the \ope{}.

It is also interesting to consider the $\mathcal{S}$-eigenvalue if the $\func{\alg{u}}{1}$-charge of the $\alg{A}_{p',p}$-\hws{} were negative.  We easily find that in this case
\begin{equation}
\phi_{h-\theta} \phi_{\theta-h} \ket{\psi} \gcreq{0} 0, \quad \text{but} \quad \phi_{h-\theta} \phi_{\theta-h} \ket{\psi} \gcreq{1} \mathcal{S} \ket{\psi}.
\end{equation}
It follows that $\mathcal{S}$ must vanish identically on any $\alg{A}_{p',p}$-\hwss{} of negative charge, hence on the corresponding Verma modules.  (We have not bracketed these calculations with $\bra{\psi}$ because we will need this conclusion to apply to \emph{singular} \hwss{} in \thmref{thmNoSingVac}.)  Note that such a bracketing shows that the first descendant $\phi_{\theta-h} \ket{\psi}$ of such an \hws{} must be singular (though this is independent of whether the \hws{} itself is singular).  In fact, an easy induction argument shows that every descendant of a \hws{} of negative charge is singular, so the corresponding irreducible module will be trivial.

\subsection{Symmetries of the Generalised Commutation Relations} \label{secRSymm}

It may seem that the algebra $\alg{A}_{p',p}$ is determined by an enormous number of \gcrs{}.  However, there is in fact a huge amount of redundancy present in these equations, which we shall now reveal.  This redundancy is exposed by two easily derived symmetries of the expressions $\func{R_{m,n}}{\gamma}$ (see \eqnref{eqnRDefined}), whose evaluation defines the \gcrs{}.  Using the form of $\func{R_{m,n}}{\gamma}$ given on the right-hand-side of \eqnref{eqnGCRGeneral}, and the binomial identity 
\begin{equation}
\binom{n}{r} - \binom{n-1}{r} = \binom{n-1}{r-1},
\end{equation}
we easily derive that
\begin{equation} \label{eqnRSymm1}
\func{R_{m,n}}{\gamma} - \func{R_{m-1,n+1}}{\gamma} = \func{R_{m,n}}{\gamma - 1}.
\end{equation}
Similarly, the left-hand-side of \eqnref{eqnGCRGeneral} makes the following symmetry evident:
\begin{equation} \label{eqnRSymm2}
\func{R_{m,n}}{\gamma} = \brac{-1}^{\gamma - 1} \func{R_{n + 2h - \gamma, m - 2h + \gamma}}{\gamma}.
\end{equation}
It is important to realise that these symmetries should be viewed as identities in $\alg{A}_{p',p}$ which make sense when applied to a state of definite monodromy charge.

\eqnref{eqnRSymm1} shows that the \gcrs{} of given order $\gamma$ contain all the information inherent in the \gcrs{} of order $\gamma - 1$ (because the latter can be derived from the former).  It follows that any \gcr{} of order $\gamma' < \gamma$ may be derived from the \gcrs{} of order $\gamma$.  This should not be surprising when we recall that $\gamma$ merely parametrises how many terms from the \ope{} (\ref{eqnOPEGeneral}) contribute to the \gcr{}.

More interestingly, if we fix the conformal dimension $m+n$ (which is of course conserved), then \eqnref{eqnRSymm1} implies that every \gcr{} of order $\gamma$ is equivalent to any other arbitrarily chosen \gcr{} of order $\gamma$, modulo those of order $\gamma - 1$.  In turn, this \gcr{} of order $\gamma - 1$ is equivalent to some arbitrarily chosen \gcr{} of order $\gamma - 1$, modulo those of order $\gamma - 2$, and so on.  Since $\func{R_{m,n}}{\gamma} = 0$ for $\gamma \leqslant 0$, we can conclude that any given \gcr{} is equivalent to a (finite) linear combination of ``basic'' \gcrs{}, one for each order parameter $\gamma$.  Furthermore, we can choose these basic relations arbitrarily.

There is one slight proviso to the above argument:  We must respect monodromy charge restrictions throughout.  Practically, this means that for a given conformal dimension $m+n$, there are in fact two disjoint classes of \gcrs{}, corresponding to the two possible monodromy charges (integral and half-integral) of the states they are to act upon.  This follows from \eqnref{eqnRSymm1} by realising that it can not be used to relate $\func{R_{m,n}}{\gamma}$ with $\func{R_{m-1/2,n+1/2}}{\gamma}$, because these would have to act on states of different monodromy charge.  A more precise version of our above conclusion is therefore that by restricting to some fixed conformal dimension $m+n$ and monodromy charge $\theta$, any given \gcr{} is equivalent to a (finite) linear combination of ``basic'' \gcrs{}, of which there is one (which we can choose arbitrarily) for each order parameter $\gamma$.

We can improve on this when $\gamma$ is \emph{even} by noting the following.  If $m - n = 2h - \gamma$, \eqnref{eqnRSymm2} implies that
\begin{equation}
\func{R_{m,n}}{\gamma} = 0 \qquad \text{($m-n = 2h - \gamma$).}
\end{equation}
Note that this may or may not be allowed by the monodromy charge of the state on which we are acting.  If it is not, we note that if $m - n = 2h - \gamma + 1$, \eqnref{eqnRSymm2} implies that $\func{R_{m-1,n+1}}{\gamma} = -\func{R_{m,n}}{\gamma}$, and substituting into \eqnref{eqnRSymm1} gives
\begin{equation}
\func{R_{m,n}}{\gamma} = \frac{1}{2} \func{R_{m,n}}{\gamma - 1} \qquad \text{($m-n = 2h - \gamma + 1$).}
\end{equation}
In either case, \eqnref{eqnRSymm1} now implies that when $\gamma$ is even, any generalised commutation relation with parameter $\gamma$ may be expressed as a linear combination of generalised commutation relations with parameter $\gamma - 1$.  In other words, the set of generalised commutation relations with even parameter $\gamma$ contains the same information as the set with parameter $\gamma - 1$.  From the point of view of a set of basic \gcrs{} (chosen arbitrarily), this means that we can drop those corresponding to $\gamma$ even.

This explains the property observed in \secref{secSEig} that computing the eigenvalue of $\mathcal{S}$ on a \hws{} of charge $\theta$ requires a \gcr{} of order $\gamma = 2 \theta + 1$ when $\theta \in \ZZ$, but only $\gamma = 2 \theta$ when $\theta \notin \ZZ$.  Essentially, $\gamma = 2 \theta + 1$ is the correct order, but when $\theta \notin \ZZ$, $2 \theta + 1$ is even, hence it is convenient to use instead the (equivalent) \gcrs{} of order $2 \theta$.

\section{Examples and Equivalences} \label{secCFTEquivs}

In this section, we investigate the structure of the extended algebras $\alg{A}_{p',p}$ with several examples.  These examples are chosen so that one can identify these extended chiral algebras with those of other familiar theories.  In other words, we provide simple (but detailed) instances of an apparent equivalence between \cfts{}.  This is followed by a careful study of what such an equivalence means at the level of chiral algebras.  We show that there are some subtleties to be addressed here, which we pin down by reconsidering what we mean by an extension of an algebra.  We then illustrate our conclusions with an interesting extended example (relegated to \appref{secM3,10}) in which a more involved \cft{} equivalence is derived.

\subsection{Examples} \label{secEx}

\subsubsection*{$\underline{\MinMod{3}{4}}$}

The simplest minimal model exhibiting a simple current is that corresponding to the Ising model, whose central charge is $c = \tfrac{1}{2}$.  The simple current $\phi = \phi_{1,3} = \phi_{2,1}$ has conformal dimension $h = \tfrac{1}{2}$, and it is well-known that the extended theory describes a free fermion.  This may be seen explicitly by substituting the first few $\func{A^{\brac{j}}}{w}$ (derived in \secref{secAlgebra}) into \eqnref{eqnOPEGeneral}, with $h = c = \tfrac{1}{2}$:
\begin{equation} \label{eqnOPEIsing}
\radord{\func{\phi}{z} \func{\phi}{w}} = \mathcal{S} \sqbrac{\frac{1}{z-w} + 2 \func{T}{w} \brac{z-w} + \func{\partial T}{w} \brac{z-w}^2 + \ldots}.
\end{equation}
The presence of the $\mathcal{S}$ in this equation (and those that follow) is not an essential complication, as it commutes with the $\phi_n$ by \eqnref{eqnSComm}, hence is a multiple of the identity on each $\alg{A}_{3,4}$-Verma module.  We may therefore treat it as a scaling factor, which we could set equal to the identity by suitably choosing the norms of the constituent $\alg{Vir}$-\hwss{} (\secref{secSEig}).  Alternatively (and equivalently), we may redefine the fermionic field as
\begin{equation}
\func{\psi}{z} = \mathcal{S}^{-1/2} \func{\phi}{z}.
\end{equation}
We note that the \ope{} (\ref{eqnOPEIsing}) implies that
\begin{equation} \label{eqnM34TDef}
\func{T}{w} = \frac{1}{2 \mathcal{S}} \normord{\func{\partial \phi}{w} \func{\phi}{w}} = \frac{1}{2} \normord{\func{\partial \psi}{w} \func{\psi}{w}},
\end{equation}
as befits a fermionic theory.  Finally, we recover the familiar anticommutation relation from the \gcr{} with $\gamma = 1$:
\begin{equation} \label{eqnM34AComm}
\phi_m \phi_n + \phi_n \phi_m \gcreq{1} \delta_{m+n,0} \mathcal{S} \qquad \Longleftrightarrow \qquad \acomm{\psi_m}{\psi_n} \gcreq{1} \delta_{m+n,0}.
\end{equation}

\subsubsection*{$\underline{\MinMod{3}{5}}$}

Now $h = \frac{3}{4}$ and $c = \frac{-3}{5}$, and we have the simplest example of a \emph{graded parafermionic} theory, $\func{\widehat{\alg{osp}}}{1 \: | \: 2}_1 / \func{\widehat{\alg{u}}}{1}$ \cite{CamGra98,JacGra02}.  This follows from the \ope{}
\begin{equation}
\radord{\func{\phi}{z} \func{\phi}{w}} = \mathcal{S} \sqbrac{\frac{1}{\brac{z-w}^{3/2}} - \frac{5}{2} \func{T}{w} \brac{z-w}^{1/2} - \ldots},
\end{equation}
and the identifications (with the notation used in \cite{CamGra98})
\begin{equation}
\func{\phi}{z} \longleftrightarrow \func{\psi_{1/2}}{z} \qquad \text{and} \qquad \func{T}{z} \longleftrightarrow \frac{-2}{5} \func{\mathcal{O}^{\brac{1/2}}}{z}.
\end{equation}
We remark that associativity (\secref{secAssoc}) forces the \ope{} of this graded parafermion $\func{\psi_{1/2}}{z}$ to involve a non-trivial operator analogous to $\mathcal{S}$ (at least when the $\func{\widehat{\alg{osp}}}{1 \: | \: 2}$-level is equal to $1$).  This was overlooked in the original treatments (but corrected in \cite{JacQua06}).

One may wonder why we do not just redefine the simple current field, as we did with $\MinMod{3}{4}$, to remove $\mathcal{S}$ from $\radord{\func{\phi}{z} \func{\phi}{w}}$.  Taking $\mathcal{S}^{1/2}$ such that $\mathcal{S}^{1/2} \phi = \ii \phi \mathcal{S}^{1/2}$, we can define $\func{\psi}{z} = e^{- \ii \pi / 4} \mathcal{S}^{-1/2} \func{\phi}{z}$ to achieve this goal.  However, we then face the problem that $\func{\psi}{z}$ satisfies the same commutativity relations as $\func{\phi}{z}$, so associativity (\secref{secAssoc}) demands that $\radord{\func{\psi}{z} \func{\psi}{w}}$ involve an $\mathcal{S}$-type operator, contradicting the fact that it was constructed so as not to.  The problem here is that it is in fact not possible to make the above redefinition because $\mathcal{S}$ does \emph{not} have such a square root\footnote{Of course $\mathcal{S}$ has \emph{many} square roots, being a self-adjoint operator on a complex (pre-)Hilbert space.  What we claim is that none of them commute with the simple current field up to a constant multiplier (necessary for the above redefinition to work).  Any prospective square root, $\mathcal{T}$ say, for which $\mathcal{T} \phi = \lambda \phi \mathcal{T}$, must satisfy $\lambda^2 = -1$, hence
\begin{equation} \label{eqnM35NoSqRt}
\mathcal{T} \ket{0} = \mathcal{T} \phi_h \phi_{-h} \ket{0} = \lambda^2 \mathcal{S} \mathcal{T} \ket{0} = \lambda^2 \mathcal{T} \ket{0} = - \mathcal{T} \ket{0} \qquad \text{(as $\phi_h \phi_{-h} \ket{0} \gcreq{1} \ket{0}$).}
\end{equation}
Obviously $\mathcal{T} \ket{0}$ cannot vanish as $\mathcal{S} \ket{0} = \ket{0}$, so our claim if proved:  No such square root $\mathcal{T}$ can exist.  Conversely, it is easy to construct such a square root when $\mathcal{S}$ commutes with $\phi$, because $\mathcal{S}$ is just a multiple of the identity on each extended Verma module.} when anticommuting with $\phi$.  

\subsubsection*{$\underline{\MinMod{3}{8}}$ and $\underline{\MinMod{4}{5}}$}

When $h = \tfrac{3}{2}$, the extended theory defines a \emph{superconformal} field theory.  We can see this explicitly by taking the \gcr{}
\begin{equation} \label{eqnGCR3h=3/2}
\phi_m \phi_n + \phi_n \phi_m \gcreq{3} \mathcal{S} \sqbrac{\binom{m + \tfrac{1}{2}}{2} \delta_{m+n,0} + \frac{3}{c} L_{m+n}},
\end{equation}
and defining
\begin{equation} \label{eqnDefG}
G_n = \sqrt{\frac{2c}{3 \mathcal{S}}} \phi_n
\end{equation}
(note that $\mathcal{S}$ and $\phi_n$ commute) to get the familiar relations
\begin{equation}
\acomm{G_m}{G_n} \gcreq{3} 2 L_{m+n} + \brac{m^2 - \tfrac{1}{4}} \delta_{m+n,0} \frac{c}{3}.
\end{equation}
The central charges of $\MinMod{3}{8}$ and $\MinMod{4}{5}$ ($\tfrac{-21}{4}$ and $\tfrac{7}{10}$) identify their extensions with the super\cfts{} $\sMinMod{2}{8}$ and $\sMinMod{3}{5}$ respectively.

\subsection{Equivalences and Algebra Isomorphisms} \label{secAlgIso}

We have just seen that the algebraic structure of the extensions of several minimal models can be identified with that of other well-known \cfts{}.  Specifically, we have based this (loose) identification on the explicit form of a \emph{finite} number of terms of the \ope{} $\radord{\func{\phi}{z} \func{\phi}{w}}$ (or on the terms of a particular \gcr{}).  However, this only guarantees that the \gcrs{} defining the extended chiral algebra have analogues under our identification when $\gamma$ is sufficiently small, hence not for arbitrary values of $\gamma$.  As might be expected, this means that it is strictly speaking \emph{incorrect} to declare that these extended minimal models are precisely the respective well-known models that we have loosely identified them with, because their chiral algebras may not be isomorphic.  We are therefore led to a consideration of precisely what we mean when we say that two \cfts{} are equivalent and equally, what we mean by saying that one is an extension of another.

Before discussing these considerations, let us remark that this concern is indeed valid, because such mismatches between chiral algebras have tangible consequences at the level of representation theory.  This is best illustrated with the example of the extended algebra of the $h = \tfrac{3}{2}$ models $\MinMod{3}{8}$ and $\MinMod{4}{5}$, considered in \secref{secEx}.  There, we identified these extended algebras with those of certain \emph{superconformal} minimal models, exhibiting a super-Virasoro algebra ($\alg{sVir}$) symmetry.  Specifically, extending $\MinMod{3}{8}$ gave (roughly speaking) $\sMinMod{2}{8}$, and extending $\MinMod{4}{5}$ gave $\sMinMod{3}{5}$.

The easiest way of seeing that the corresponding chiral algebras are \emph{not} isomorphic is to note that the $\alg{A}_{3,8}$ and $\alg{A}_{4,5}$-Verma modules are irreducible (as we shall prove in \secref{secGenSingVect}), whereas it is well-known that those of the corresponding super-Virasoro algebras are not (they contain non-trivial singular vectors).  As the definitions of \hws{} and Verma module for these two chiral algebras are compatible\footnote{We refer to \cite[Sec.\ 5.4]{RidSU206} for an example where they are not.}, the notion of irreducibility should be preserved by any algebra isomorphism.  We are therefore forced to conclude that $\alg{A}_{3,8}$ and $\alg{A}_{4,5}$ are \emph{not} actually isomorphic to (the \uea{} of) the corresponding super-Virasoro algebras, $\alg{sVir}_{2,8}$ and $\alg{sVir}_{3,5}$, respectively.

The identification made in \secref{secEx} (which we have just questioned) was limited to \eqnref{eqnGCR3h=3/2}, the \gcrs{} with $\gamma = 3$.  The results of \secref{secRSymm} demonstrate that these \gcrs{} imply those with $\gamma < 3$, but not necessarily those with $\gamma \geqslant 5$.  Our non-isomorphism result therefore requires that there exist further identities satisfied in $\alg{A}_{3,8}$ and $\alg{A}_{4,5}$, which have no counterpart in the super-Virasoro algebra (and are responsible for the irreducibility of the Verma modules of the extended theory).

For example, we could consider the \gcr{} with $\gamma = 4$, which receives a contribution from the first regular term in the \ope{} $\radord{\func{\phi}{z} \func{\phi}{w}}$.  At the level of fields, we have (using \eqnref{eqnDefG})
\begin{equation}
\normord{\func{\phi}{w} \func{\phi}{w}} = \func{A^{\brac{4}}}{w} = \frac{3 \mathcal{S}}{2 c} \func{\partial T}{w} \qquad \Rightarrow \qquad \normord{\func{G}{w} \func{G}{w}} = \func{\partial T}{w}.
\end{equation}
The latter relation is of course standard for super\cfts{}, and reflects the ($\gamma = 3$) anticommutation relation
\begin{equation}
\acomm{G_{-3/2}}{G_{-3/2}} = 2 L_{-3}.
\end{equation}
(This illustrates the fact that when $\gamma$ is even, the \gcrs{} of order $\gamma$ may be obtained from those of order $\gamma - 1$.)  The \gcrs{} with $\gamma = 5$ however imply the identity
\begin{equation}
\normord{\func{\partial G}{w} \func{G}{w}} = \frac{17}{5c+22} \normord{\func{T}{w} \func{T}{w}} + \frac{3 \brac{c+1}}{2 \brac{5c+22}} \func{\partial^2 T}{w},
\end{equation}
which is not a generic identity in super\cft{}.  Indeed, we will see in \secref{secExamples} that such a relation cannot be satisfied identically, because the $\gamma = 5$ \gcrs{} are sufficient to prove the vanishing of the first singular vector in the extended algebra vacuum Verma module, but this vector does not vanish identically in the super-Virasoro vacuum Verma module.

This example is typical of the general situation:  Comparing a finite number of terms in the defining \opes{} (or \gcrs{}) is not generally enough to demonstrate an isomorphism of chiral algebras (even if all singular terms are included).  To make this point precise, we introduce a sequence of extended chiral algebras $\alg{A}_{p',p}^{\brac{\gamma}}$, $\gamma \in \ZZ$, in which only the \gcrs{} of order $\gamma$ are imposed.  The results of \secref{secRSymm} allow us to depict this sequence as follows:
\begin{equation}
\alg{F}_{p',p} \rightarrow \cdots \rightarrow \alg{A}_{p',p}^{\brac{-1}} \xrightarrow{\cong} \alg{A}_{p',p}^{\brac{0}} \rightarrow \alg{A}_{p',p}^{\brac{1}} \xrightarrow{\cong} \alg{A}_{p',p}^{\brac{2}} \rightarrow \alg{A}_{p',p}^{\brac{3}} \xrightarrow{\cong} \alg{A}_{p',p}^{\brac{4}} \rightarrow \alg{A}_{p',p}^{\brac{5}} \rightarrow \cdots \rightarrow \alg{A}_{p',p}.
\end{equation}
Here, $\alg{F}_{p',p}$ denotes the free algebra generated by the $\phi_n$ modes, and the arrow from $\alg{A}_{p',p}^{\brac{\gamma - 1}}$ to $\alg{A}_{p',p}^{\brac{\gamma}}$ represents the quotient map obtained by factoring out the (two-sided) ideal generated by the \gcrs{} of order $\gamma$.  The ``fully'' extended algebra $\alg{A}_{p',p}$ is then the \emph{direct limit} of these ``partially'' extended algebras:
\begin{equation}
\alg{A}_{p',p} = \varinjlim \alg{A}_{p',p}^{\brac{\gamma}}.
\end{equation}

In this formalism, the precise version of the correspondence between $\alg{A}_{3,8}$, $\alg{A}_{4,5}$ and the appropriate super-Virasoro algebras is as follows\footnote{We remark that the algebra isomorphisms constructed in \cite[Sec.\ 4]{RidSU206} must also be understood in this sense.  Specifically, if we denote the extended chiral algebra of the level-$k$ $\func{\group{SU}}{2}$ Wess-Zumino-Witten model by $\widehat{\alg{A}}_k$, then the precise isomorphisms proved there are
\begin{equation}
\widehat{\alg{A}}_2^{\brac{3}} \cong \alg{A}_{3,4}^{\brac{3}} \otimes \alg{A}_{3,4}^{\brac{3}} \otimes \alg{A}_{3,4}^{\brac{3}} \qquad \text{and} \qquad \widehat{\alg{A}}_4^{\brac{2}} \cong \uealg{\func{\widehat{\alg{sl}}}{3}_1} \ncong \widehat{\alg{A}}_4.
\end{equation}
We expect that the first of these isomorphisms can be extended to all $\gamma \geqslant 3$ (hence for the fully extended algebras).  The presence of non-trivial $\func{\widehat{\alg{sl}}}{3}_1$-singular vectors implies that such an extension is not possible in the second case (hence the non-isomorphism indicated).}:
\begin{equation} \label{Isosh=3/2}
\alg{A}_{3,8}^{\brac{3}} \cong \uealg{\alg{sVir}_{2,8}} \ncong \alg{A}_{3,8}^{\brac{5}} \qquad \text{and} \qquad \alg{A}_{4,5}^{\brac{3}} \cong \uealg{\alg{sVir}_{3,5}} \ncong \alg{A}_{4,5}^{\brac{5}}.
\end{equation}
The compatibility of the respective definitions of a \hws{} imply that these isomorphic chiral algebras have isomorphic Verma modules, and more importantly, isomorphic irreducible \hwms{}.  In general then, a simple current defines an infinite family of extended chiral algebras, parametrised by the order ($\gamma$) of the defining \gcrs{}.  In this sense, $\uealg{\alg{sVir}_{2,8}}$ is \emph{an} extension of $\uealg{\alg{Vir}_{3,8}}$, but not a ``maximal'' extension (in the obvious sense).

Physically however, it is not the Verma modules which are fundamental, but the corresponding irreducible modules.  We might expect that different extensions (defined by different $\gamma$) would have isomorphic irreducible representations, and this is indeed true under one important proviso.  This follows from the fact that it is immaterial to the construction of the irreducible modules from the Verma modules whether the singular vectors vanish identically or are quotiented out.  The proviso follows from the fact that the algebraic structure of the extensions must be sufficient to be able to compute whether a given vector is singular or not.  The general, but fundamental, requirement of actually being able to compute with a given symmetry algebra leads to quite non-trivial bounds on $\gamma$.  For example, we have seen in \secref{secSEig} that computing $\mathcal{S}$-eigenvalues requires $\gamma > 2h$.

A full analysis of this proviso will not be required here, because we only consider extensions in which we know \emph{a priori} that computation is possible.  The upshot is then that the isomorphisms and non-isomorphisms of (\ref{Isosh=3/2}) still imply the corresponding isomorphisms between the irreducible $\alg{A}_{3,8}$ and $\alg{A}_{4,5}$-modules and the irreducible super-Virasoro modules (though not between the corresponding Verma modules).  These isomorphisms therefore detail \emph{explicitly} the \cft{} equivalences
\begin{equation}
\MinMod{3}{8} \equiv \sMinMod{2}{8} \qquad \text{and} \qquad \MinMod{4}{5} \equiv \sMinMod{3}{5},
\end{equation}
and imply the corresponding character identities.

We conclude this discussion by mentioning the other equivalences noted in \secref{secEx}, involving the theories describing the free fermion and the (level $1$) graded parafermion.  If we denote their chiral algebras by $\alg{F}$ and $\alg{G}$ respectively, then the algebra isomorphisms described there take the form
\begin{equation}
\alg{A}_{3,4}^{\brac{3}} \cong \alg{F} \qquad \text{and} \qquad \alg{A}_{3,5}^{\brac{3}} \cong \alg{G}.
\end{equation}
Here, we define $\alg{F}$ as the associative algebra defined by the anticommutation relations of \eqnref{eqnM34AComm} and the relation
\begin{equation}
L_n = \frac{-1}{2} \sum_m \brac{m + \frac{1}{2}} \normord{\phi_m \phi_{n-m}},
\end{equation}
corresponding to \eqnref{eqnM34TDef} (and necessary for conformal symmetry).  Similarly, we define $\alg{G}$ as the associative algebra defined by the \gcrs{} given in \cite{CamGra98}.  We will see in \secref{secExamples} that for $\alg{A}_{3,4}$, $\gamma = 3$ is sufficient to prove that all singular vectors vanish.  Therefore, even if there were further non-trivial algebraic relations (corresponding to $\gamma \geqslant 5$ for example), these relations would have to act trivially on every representation (irreducible \emph{or} Verma).

\section{Singular Vectors} \label{secSingVect}

Recall from \secref{secVerma} that there is an isometric surjection (\ref{Surjection}) mapping the (direct sum of) the two constituent $\alg{Vir}$-Verma modules onto the corresponding $\alg{A}_{p',p}$-Verma module.  Furthermore, this is a homomorphism of $\alg{Vir}$-modules.  The natural question to ask then is which vectors get mapped to zero.  In other words, what is the kernel of this homomorphism.  As our homomorphism is norm-preserving, we immediately see that this kernel can only consist of null vectors.  Clarifying our original question slightly, we are led to ask whether \emph{all} the $\alg{Vir}$-null vectors are mapped to zero, that is, is the kernel precisely the set of null vectors?  A negative answer to this question necessarily requires that there exist non-trivial singular vectors in the $\alg{A}_{p',p}$-Verma module.

Now, the irreducible $\alg{Vir}$-modules comprising the minimal models are well-known to be quotients of the corresponding Verma modules by a submodule generated by two\footnote{In the vacuum Verma module $\VirVerMod{p',p}{1,1}$, the first of these is $L_{-1} \ket{0}$, which is usually taken to vanish identically.  Physically, this reflects the requirement that the (chiral) vacuum be invariant under the (chiral) global conformal transformations generated by $L_1$, $L_0$ and $L_{-1}$ --- that is, under the (chiral) conformal group.  Mathematically, this may be derived as a simple consequence of the state-field correspondence:  $L_{-1} \ket{0}$ corresponds to the derivative of the identity field.} singular vectors, which we refer to as the \emph{principal} singular vectors.  The explicit form of these singular vectors is only known in special cases \cite{BenDeg88}, but their grades follow easily from Kac's determinant formula.  Specifically, the $\alg{Vir}$-Verma module $\VirVerMod{p',p}{r,s}$ has principal singular vectors at grades $rs$ and $\brac{p' - r} \brac{p - s}$.

We will restrict ourselves to considering the images of the principal singular vectors in the vacuum $\alg{A}_{p',p}$-Verma module $\ExtVerMod{p',p}{1,1}$.  There are three to consider, at grades $p'-1$ and $p-1$, corresponding to the simple current $\alg{Vir}$-module, and at grade $\brac{p' - 1} \brac{p - 1}$, corresponding to the vacuum $\alg{Vir}$-module.  We will first investigate the explicit form of these singular vectors in the simplest minimal models, before lifting our conclusions to the general case.  For clarity, we will usually restrict attention to the principal singular vector of lowest grade.

\subsection{Examples} \label{secExamples}

\subsubsection*{$\underline{\MinMod{3}{4}}$}

We recall the well-known fact that the irreducible modules (Fock spaces) of the extended algebra are freely generated, that is, no null vectors are encountered in their construction.  Indeed, the first (principal) singular vector of the extended vacuum module is
\begin{equation}
\begin{split}
\brac{L_{-2} - \frac{3}{4} L_{-1}^2} \ket{\phi} &= \brac{L_{-2} - \frac{3}{4} L_{-1}^2} \phi_{-1/2} \ket{0} = \phi_{-1/2} L_{-2} \ket{0} \\
&\gcreq{3} \frac{1}{2} \phi_{-1/2} \phi_{-3/2} \phi_{-1/2} \ket{0},
\end{split}
\end{equation}
which clearly vanishes due to the anticommutation relations (\ref{eqnM34AComm}).  Similarly, the second principal singular vector is
\begin{equation}
\begin{split}
\brac{L_{-3} - 4 L_{-2} L_{-1} + \frac{4}{3} L_{-1}^3} \phi_{-1/2} \ket{0} &= \brac{L_{-3} \phi_{-1/2} - 4 L_{-2} \phi_{-3/2} + 8 \phi_{-7/2}} \ket{0} \\
&= \brac{\phi_{-1/2} L_{-3} - 4 \phi_{-3/2} L_{-2}} \ket{0} \\
&\gcreq{3} \brac{\frac{1}{4} \phi_{-1/2} \phi_{-5/2} \phi_{-1/2} - 2 \phi_{-3/2} \phi_{-3/2} \phi_{-1/2}} \ket{0} \\
&= 0.
\end{split}
\end{equation}
Finally, a direct computation using the $\gamma = 3$ generalised commutation relation shows that the last singular vector
\begin{equation} \label{eqnM34VacSing}
\brac{L_{-6} + \frac{22}{9} L_{-4} L_{-2} - \frac{31}{36} L_{-3}^2 - \frac{16}{27} L_{-2}^3} \ket{0}
\end{equation}
may be expressed as a linear combination of the vectors $\phi_{-11/2} \phi_{-1/2} \ket{0}$, $\phi_{-9/2} \phi_{-3/2} \ket{0}$, and $\phi_{-7/2} \phi_{-5/2} \ket{0}$, whose coefficients exactly cancel.

To summarise, the singular vectors of the extended Verma module $\ExtVerMod{3,4}{1,1}$ all vanish identically.  This Verma module is thus irreducible.  The same can easily be verified for the remaining $\alg{A}_{3,4}$-Verma module $\ExtVerMod{3,4}{1,2}$, whose highest weight state has conformal dimension $\tfrac{1}{16}$.

\subsubsection*{$\underline{\MinMod{3}{5}}$}

It is not hard to show that the first (principal) singular vector may again be shown to vanish identically:
\begin{align}
\brac{L_{-2} - \frac{3}{5} L_{-1}^2} \phi_{-3/4} \ket{0} &\gcreq{3} \brac{\frac{-4}{5} \phi_{-7/4} \phi_{-1/4} \mathcal{S}^{-1} \phi_{-3/4} - \frac{6}{5} \phi_{-11/4} \phi_{3/4} \mathcal{S}^{-1} \phi_{-3/4} - \frac{6}{5} \phi_{-11/4}} \ket{0} \notag \\
&= \brac{\frac{4}{5} \phi_{-7/4} \phi_{-1/4} \phi_{-3/4} + \frac{6}{5} \phi_{-11/4} \phi_{3/4} \phi_{-3/4} - \frac{6}{5} \phi_{-11/4}} \ket{0} \\
&= \brac{\frac{6}{5} \phi_{-11/4} - \frac{6}{5} \phi_{-11/4}} \ket{0} = 0. \notag
\end{align}
In the third equality here, we have made use of the simple relations
\begin{equation}
\phi_{h-1} \phi_{-h} \ket{0} \gcreq{1} 0 \qquad \text{and} \qquad \phi_h \phi_{-h} \ket{0} \gcreq{1} \ket{0},
\end{equation}
which apply quite generally and which will be used without comment in the future.  We remark that to obtain this vanishing result, we must remember that $\mathcal{S}$ anticommutes with the $\phi_n$ and leaves $\ket{0}$ invariant.  It should be clear that if there were no $\mathcal{S}$-operator, then the final result above would be $\tfrac{-12}{5} \phi_{-11/4} \ket{0}$ rather than zero, implying that
\begin{equation}
\brac{L_{-2} + \frac{3}{5} L_{-1}^2} \ket{\phi} = 0
\end{equation}
(which contradicts the fact that this linear combination is not even null).

The second principal singular vector also vanishes identically, although the demonstration of this is deferred to \appref{appM35}.  This computation captures the essential complications of such a verification for the second primary singular vector.  The vanishing of the Virasoro vacuum singular vector (at grade $8$) may be demonstrated similarly, but this and the corresponding computations for the other $\alg{A}_{3,5}$-Verma module will be omitted.

\subsubsection*{$\underline{\MinMod{3}{7}}$}

We have $h = \frac{5}{4}$ and $c = \frac{-25}{7}$.  The first (principal) singular vector is
\begin{equation}
\begin{split}
\brac{L_{-2} - \frac{3}{7} L_{-1}^2} \phi_{-5/4} \ket{0} &= \brac{\phi_{-5/4} L_{-2} + \frac{3}{4} \phi_{-13/4} - \frac{6}{7} \phi_{-13/4}} \ket{0} \\
&\gcreq{3} \brac{\frac{-10}{7} \phi_{-5/4} \phi_{-3/4} \phi_{-5/4} - \frac{3}{28} \phi_{-13/4}} \ket{0} \\
&\gcreq{3} \brac{\frac{-10}{7} \sqbrac{\frac{7}{20} L_{-2} - \frac{3}{8} \phi_{-13/4} \phi_{5/4}} \phi_{-5/4} - \frac{3}{28} \phi_{-13/4}} \ket{0} \\
&= \frac{-1}{2} \brac{L_{-2} - \frac{3}{7} L_{-1}^2} \phi_{-5/4} \ket{0},
\end{split}
\end{equation}
and so it again vanishes.  This calculation typifies the general procedure:  The Virasoro modes are commuted to the right until they act on the vacuum, then the $\gamma = 3$ \gcr{} is used to write them in terms of $\phi$-modes.  We then apply a suitable \gcr{} to the two leftmost $\phi$-modes in order to re-express them in terms of Virasoro modes, finally getting back a multiple of the singular vector.

\subsubsection*{$\underline{\MinMod{3}{8}}$}

We consider the first singular vector of $\MinMod{3}{8}$ ($c = \frac{-21}{4}$), and apply the procedure outlined above.  Commuting and using $\gamma = 3$ gives
\begin{equation}
\brac{L_{-2} - \frac{3}{8} L_{-1}^2} \ket{\phi} \gcreq{3} \brac{\frac{-7}{4} \phi_{-3/2} \phi_{-1/2} - \frac{1}{8} L_{-1}^2} \ket{\phi}.
\end{equation}
We now apply $\gamma = 5$ to $\phi_{-3/2} \phi_{-1/2}$ and tidy up, getting
\begin{equation}
\brac{L_{-2} - \frac{3}{8} L_{-1}^2} \ket{\phi} \gcreq{5} -6 \brac{L_{-2} - \frac{3}{8} L_{-1}^2} \ket{\phi},
\end{equation}
hence the (by now) expected vanishing.

\subsubsection*{$\underline{\MinMod{4}{5}}$}

If we try to apply this procedure to the $\MinMod{4}{5}$ vector $\brac{L_{-2} - \frac{3}{8} L_{-1}^2} \ket{\phi}$ ($c = \frac{7}{10}$), we find that we end up with precisely this vector again, rather than a non-trivial multiple of it:
\begin{equation}
\brac{L_{-2} - \frac{3}{8} L_{-1}^2} \ket{\phi} \gcreq{3} \brac{\frac{7}{30} \phi_{-3/2} \phi_{-1/2} - \frac{1}{8} L_{-1}^2} \ket{\phi} \gcreq{5} \brac{L_{-2} - \frac{3}{8} L_{-1}^2} \ket{\phi}.
\end{equation}
Of course, this vector is not null in the $\MinMod{4}{5}$ model, so it should not be surprising that it does not vanish identically.  The first singular vector is in fact
\begin{equation}
\brac{L_{-3} - \frac{4}{3} L_{-2} L_{-1} + \frac{4}{15} L_{-1}^3} \ket{\phi},
\end{equation}
and our procedure with $\gamma = 3$, $5$, and then $5$ again, proves that it indeed vanishes identically.

\subsection{The General Case} \label{secGenSingVect}

Based on these computations, it seems reasonable to conjecture that all the Virasoro singular vectors vanish identically when mapped into the appropriate $\alg{A}_{p',p}$-Verma module.  A proof is presented below for the vacuum module.  Unfortunately, the same argument  cannot be extended to all $\alg{A}_{p',p}$-Verma modules.  In particular, it breaks down completely when applied to modules of charge $h$.  However, after presenting this proof, we will discuss why the vanishing of the singular vectors in the $\alg{A}_{p',p}$-vacuum Verma module in fact implies the corresponding result for the other Verma modules.  We mention in passing that the method established in \secref{secExamples} to demonstrate the vanishing of the singular vectors requires using the generalised commutation relations with $\gamma = 4h - 1$ (or, equivalently, $4h - 2$ when $4h$ is odd).  These clearly receive contributions from many regular terms in the \ope{} (\ref{eqnOPEGeneral}).

\begin{theorem} \label{thmNoSingVac}
The vacuum $\alg{A}_{p',p}$-Verma module $\ExtVerMod{p',p}{1,1}$ of the $\MinMod{p'}{p}$ model with $p > p' > 2$ has no singular vectors, hence is irreducible.
\end{theorem}

\noindent In \cite[Thm.\ 5.3]{RidSU206}, we have established the analogous result for the $\func{\group{SU}}{2}$ Wess-Zumino-Witten models (and all modules) by direct, though highly non-trivial, computation.  Such a direct proof is not possible in the present context as we do not have as detailed a knowledge of the explicit form of the principal Virasoro singular vectors.  The method of proof we give below is therefore indirect, being essentially a proof by contradiction.  We remark that this contradiction argument is in fact quite general and should serve as a prototype for future generalisations.

\begin{figure}
\begin{center}
\includegraphics[width=13cm]{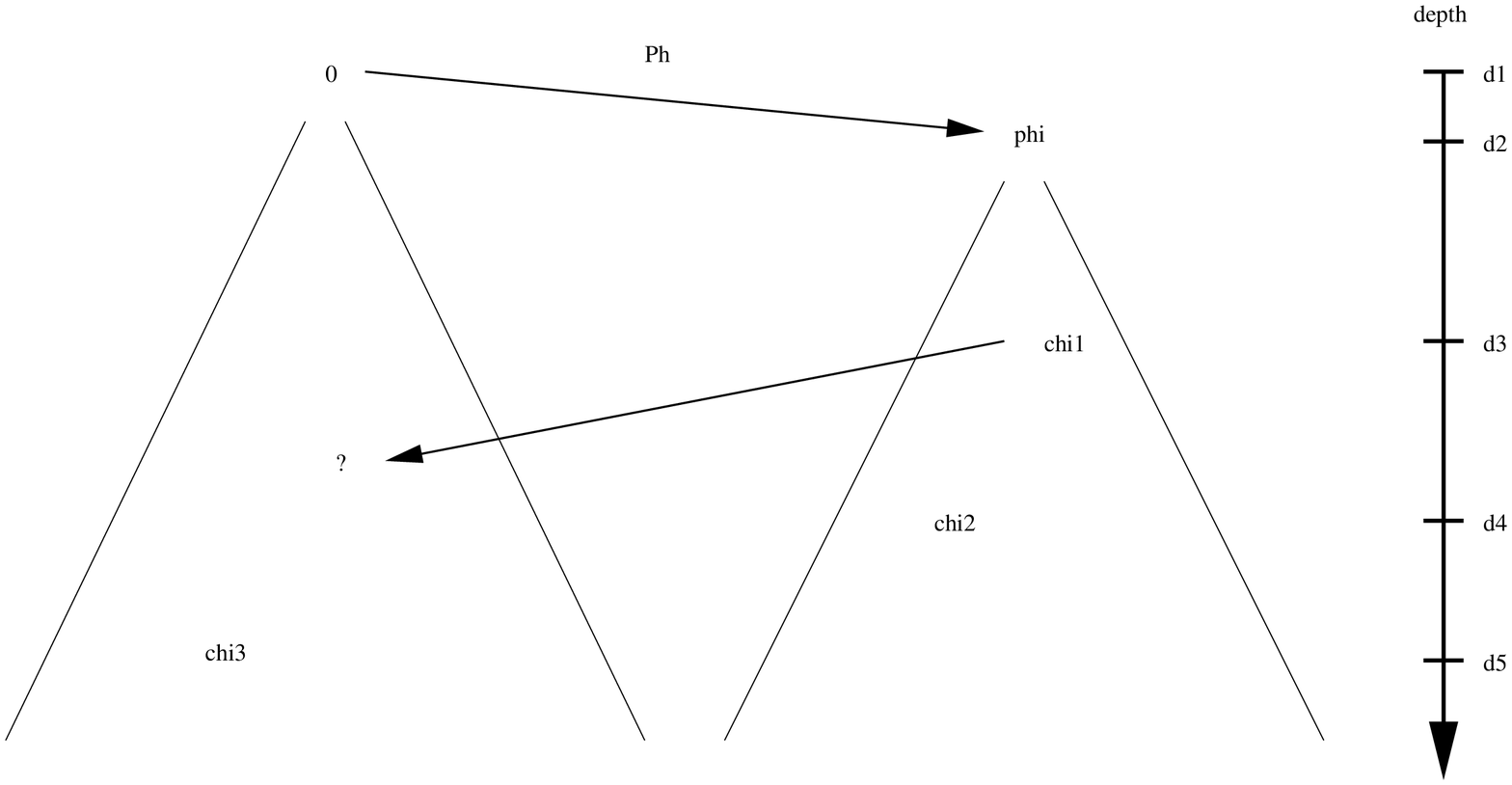}
\caption{A pictorial representation of the vacuum $\alg{A}_{p',p}$-Verma module $\ExtVerMod{p',p}{1,1}$, illustrating the mechanism behind the proof of \thmref{thmNoSingVac}.  The states $\ket{\chi_i}$, $i = 1, 2, 3$, are the principal Virasoro singular vectors, and $\ket{?}$ is the purported singular descendant whose non-existence gives the contradiction.} \label{figNoSingVacProof}
\end{center}
\end{figure}

Before giving the formal proof of \thmref{thmNoSingVac}, let us outline the simple idea behind it.  The vacuum $\alg{A}_{p',p}$-Verma module consists of two Virasoro modules (of unknown character).  If one of these modules contained a non-vanishing singular vector, then it would have to possess a non-vanishing singular $\alg{A}_{p',p}$-descendant \emph{in the other Virasoro module}.  By comparing the depth (conformal dimension) of this supposedly singular descendant with the depths of the known singular vectors of this other module, we derive a contradiction, showing that there could not have been such a non-vanishing singular vector to start with.  We illustrate this idea schematically in \figref{figNoSingVacProof}.

\psfrag{0}[][]{$\ket{0}$}
\psfrag{phi}[][]{$\ket{\phi}$}
\psfrag{chi1}[][]{$\ket{\chi_1}$}
\psfrag{chi2}[][]{$\ket{\chi_2}$}
\psfrag{chi3}[][]{$\ket{\chi_3}$}
\psfrag{?}[][]{$\ket{?}$}
\psfrag{Ph}[][]{$\phi_{-h}$}
\psfrag{depth}[][]{Depth}
\psfrag{d1}{$0$}
\psfrag{d2}{$h$}
\psfrag{d3}{$h+p'-1$}
\psfrag{d4}{$h+p-1$}
\psfrag{d5}{$\brac{p'-1} \brac{p-1}$}

\begin{proof}
We argue by contradiction.  Let us begin with the first primitive singular vector $\ket{\chi_1}$ of the $\alg{Vir}$-Verma module $\VirVerMod{p',p}{p'-1,1}$ at grade $p'-1$.  In $\ExtVerMod{p',p}{1,1}$, $\ket{\chi_1}$ is a $\alg{Vir}$-\hws{} of conformal dimension $p' - 1 + h$.  As there can be no null vectors of conformal dimension lower than this in $\ExtVerMod{p',p}{1,1}$, it follows that $\ket{\chi_1}$ must in fact be an $\alg{A}_{p',p}$-\hws{}.
 Suppose that the singular vector $\ket{\chi_1}$ does \emph{not} vanish identically (although its norm does of course).  It must then have a $\func{\alg{u}}{1}$-charge $\theta$ (not exceeding $h$).  In addition, $\ket{\chi_1}$ is an eigenstate of $\mathcal{S}$ with eigenvalue $\brac{-1}^{4h} \neq 0$, from which we conclude that $\theta$ cannot be negative (\secref{secSEig}).  As the first descendant of $\ket{\chi_1}$ is obtained by acting with $\phi_{\theta - h}$, it has conformal dimension $p' - 1 + 2h - \theta$.  This descendant is non-zero by definition, and its norm obviously vanishes:
\begin{equation}
\bracket{\chi_1}{\phi_{h - \theta} \phi_{\theta - h}}{\chi_1} \propto \braket{ \chi_1}{\chi_1} = 0.
\end{equation}
On the other hand, $\phi_{\theta - h} \ket{\chi_1}$ is a state in $\VirVerMod{p',p}{1,1}$ which is not a descendant of $L_{-1} \ket{0}$ (as this has been taken to vanish identically).  It is therefore either the Virasoro vacuum singular vector, or a (Virasoro) descendant thereof.  The desired contradiction is now obtained (as $p \geqslant 4$) because the first singular element $\ket{\chi_3}$ of $\VirVerMod{p',p}{1,1}$ (hence the first state with zero norm that has not yet been proven to vanish identically) has conformal dimension $\brac{p'-1} \brac{p-1}$, and
\begin{equation}
p' - 1 + 2h - \theta \geqslant \brac{p' - 1} \brac{p-1} \qquad \Rightarrow \qquad p \leqslant 2 - \frac{2 \theta}{p'} \leqslant 2.
\end{equation}
The singular vector $\ket{\chi_1}$ therefore vanishes identically.  Clearly its Virasoro descendants will then also vanish identically.

The next primitive singular vector is $\ket{\chi_2} \in \VirVerMod{p',p}{p'-1,1}$ at grade $p-1$.  Its conformal dimension is then $p - 1 + h$.  Since we have just shown that there are no non-vanishing null vectors of lower conformal dimension in $\ExtVerMod{p',p}{1,1}$, $\ket{\chi_2}$ must also be an $\alg{A}_{p',p}$-\hws{} of charge $0 \leqslant \theta \leqslant h$.  If non-vanishing, it would necessarily have a non-vanishing descendant of conformal dimension $p - 1 + 2h - \theta$, which is smaller than $\brac{p'-1} \brac{p-1}$ unless $p' \leqslant 2 - \frac{2 \theta}{p} \leqslant 2$, another contradiction (as $p' \geqslant 3$).  So, $\ket{\chi_2}$ and its Virasoro descendants also vanish identically.

Finally, the primitive singular vector $\ket{\chi_3} \in \VirVerMod{p',p}{1,1}$ at grade $\brac{p'-1} \brac{p-1}$ should have, if non-vanishing itself, a non-vanishing singular descendant in $\VirVerMod{p',p}{p'-1,1}$.  But we have seen that all the singular vectors in $\VirVerMod{p',p}{p'-1,1}$ vanish identically, hence there is no such descendant.  This last contradiction proves the result.
\end{proof}

\noindent Abstractly, this result may be restated as follows:  Every Virasoro singular vector, hence every null vector, in $\VirVerMod{p',p}{1,1}$ and $\VirVerMod{p',p}{p'-1,1}$ is mapped to zero by the surjection (\ref{Surjection}).  Thus, we have completely characterised the kernel of this map.  This is summarised in the following corollary.

\begin{corollary} \label{corIrreducible}
When $r = s = 1$, the kernel of the surjection (\ref{Surjection}) is precisely the submodule of Virasoro null vectors, hence
\begin{equation} \label{Isomorphism}
\VirIrrMod{p',p}{1,1} \oplus \VirIrrMod{p',p}{p'-1,1} \cong \ExtVerMod{p',p}{1,1} \cong \ExtIrrMod{p',p}{1,1},
\end{equation}
where $\VirIrrMod{}{}$ and $\ExtIrrMod{}{}$ denote the \emph{irreducible} $\alg{Vir}_{p',p}$ and $\alg{A}_{p',p}$-\hwms{} (respectively).
\end{corollary}

We consider now the argument behind the proof of \thmref{thmNoSingVac}, as it applies to the other $\alg{A}_{p',p}$-modules of the theory.  The problem with this argument is that it is easy to see that it cannot work for all modules.  The most extreme example illustrating this is an $\alg{A}_{p',p}$-module whose \hws{} has charge $h$.  Such modules are composed of two identical $\alg{Vir}$-modules, connected by the mode $\phi_0$.  Evidently, a singular vector in one of these $\alg{Vir}$-modules has a counterpart in the other at the same grade, so establishing our conclusion the same way would amount to justifying that the charge of the singular vector is greater than $h$ (contradicting it being an $\alg{A}_{p',p}$-\hws{}).

Instead, we recall that in a rational \cft{} the vacuum singular vector is supposed to determine which \hwms{} can be consistently added to the theory.  In other words, it controls the spectrum of the theory.  Specifically, a non-trivial singular vector $\ket{\chi}$ of the vacuum module corresponds to a null field $\func{\chi}{z}$ under the state-field correspondence.  The zero-mode $\chi_0$ of the null field then selects the allowed \hwss{} through the simple requirement that
\begin{equation}
\chi_0 \ket{\psi} = 0.
\end{equation}
It is worth emphasising that this is a consistency requirement that all \cfts{} must satisfy.  It has been checked explicitly for the $\MinMod{2}{p}$ models in \cite{FeiAnn92}, where it is claimed that the result for general minimal models was proven in \cite{FeiCoh88}.

This begs the question:  Why do the singular vectors of the other \hwms{} of the theory not further restrict the spectrum?  The answer appears to be:  Because these other singular vectors may all be obtained from the vacuum singular vector.  More precisely, we have:
\begin{claim}
If $\ket{\chi}$ is the $\MinMod{p'}{p}$ vacuum singular vector at grade $\brac{p'-1} \brac{p-1}$, and $\ket{\zeta_{r,s}}$ is the \emph{first} principal singular vector of the Verma module headed by $\ket{\psi_{r,s}}$ (with $r$ and $s$ chosen so that $rs < \brac{p' - r} \brac{p - s}$), then
\begin{equation}
\ket{\zeta_{r,s}} = \chi_{-rs} \ket{\psi_{r,s}},
\end{equation}
up to an arbitrary normalisation.  The second principal singular vector is then a linear combination of $\chi_{- \brac{p' - r} \brac{p - s}} \ket{\psi_{r,s}}$ and Virasoro descendants of $\ket{\zeta_{r,s}}$.
\end{claim}
\noindent Obviously, $\chi_{-rs} \ket{\psi_{r,s}}$ is null, hence proportional to $\ket{\zeta_{r,s}}$, so our Claim is just that this proportionality constant is non-zero.  Note that $\chi_n \ket{\psi_{r,s}}$ vanishes identically for all $n < rs$.

This is not hard to verify in simple cases (for instance, $\MinMod{2}{5}$ and $\MinMod{3}{4}$).  Indeed, it follows easily from \cite[Thm.\ 3.6]{FeiAnn92} that it is true for all $\MinMod{2}{p}$ models.  However, we are not aware of proofs of this Claim for general minimal models, even though we are certain of its truth.  We shall therefore assume that it is valid in what follows.

The power of this Claim should be evident.  By proving that the vacuum singular vector identically vanishes in the extended theory (itself a consequence of the vanishing of the singular vectors of the simple current module), we obtain the vanishing of the null field $\func{\chi}{z}$, and hence the vanishing of each mode $\chi_n$.  If the singular vectors of every (allowed) Virasoro module are induced by these modes acting on the relevant \hwss{}, then they must also vanish identically in the extended theory.  The implication of \corref{corIrreducible} is therefore the following strengthening:
\begin{corollary} \label{corIrreducibleGen}
Assuming the above Claim, 
\begin{equation}
\VirIrrMod{p',p}{r,s} \oplus \VirIrrMod{p',p}{p'-r,s} \cong \ExtVerMod{p',p}{r,s} \cong \ExtIrrMod{p',p}{r,s},
\end{equation}
for all $1 \leqslant r \leqslant p'-1$, $1 \leqslant s \leqslant p-1$.
\end{corollary}
\noindent In other words, every such $\alg{A}_{p',p}$-Verma module is irreducible, and decomposes under $\alg{Vir}_{p',p}$ into the direct sum of two irreducible modules.  This is our main result.

Let us conclude this section with a final remark.  From the point of view of the algebra $\alg{A}_{p',p}$, it is not natural to leave the central role in determining the spectrum to the Virasoro vacuum singular vector.  It would be more natural to have a picture in which the two principal singular vectors of the simple current module would themselves control the whole spectrum (after all, it is the \ope{} of the simple current with itself that is supposed to contain all the information of the theory).  We will present computations supporting this expectation in the next section.

\section{Some Simple Applications} \label{secApps}

The absence of the singular vectors in the $\alg{A}_{p',p}$-modules which has just been demonstrated has immediate structural consequences.  In particular, the first principal singular vector of $\phi$ vanishing identically implies powerful recurrence relations for the fields $\func{A^{\brac{j}}}{w}$ appearing in the \ope{} (\ref{eqnOPEGeneral}) of $\func{\phi}{z}$ with itself.  These relations provide a fundamental calculational tool for computing within these extended algebras.  We illustrate this by showing how they may be used to efficiently determine the $\mathcal{S}$-eigenvalues (proceeding directly as in \secref{secSEig} rapidly becomes cumbersome when the monodromy charges are large).  As fundamental computational tools, it is important to determine the applicability of these recursion relations.  We will see that on occasion a given recursion relation may fail to determine a small (finite) number of the $\func{A^{\brac{j}}}{w}$, and isolate this phenomenon precisely through an amusing exercise in number theory.  We conclude the section with a brief discussion on the role of the singular vectors of the extended vacuum module in restricting the spectrum of the theory.

\subsection{Recursion Relations for the $\ket{A^{\brac{j}}}$} \label{secOPERecursive}

We illustrate the derivation of these recursion relations for the $\MinMod{3}{p}$ models, for which $h = \tfrac{1}{4} \brac{p-2}$.  The first principal singular vector of the simple current module is 
\begin{equation} \label{eqnM3pFirstSV}
\brac{L_{-2} - \frac{3}{2 \brac{2h + 1}} L_{-1}^2} \ket{\phi},
\end{equation}
so its vanishing in $\ExtVerMod{3,p}{2,1}$ and the usual commutation relations give
\begin{equation}
\begin{split}
0 &= \phi_{h-j+2} \brac{L_{-2} - \frac{3}{2 \brac{2h + 1}} L_{-1}^2} \ket{\phi} \\
&= \sqbrac{\brac{L_{-2} - \frac{3}{2 \brac{2h + 1}} L_{-1}^2} \phi_{h-j+2} - \frac{3 \brac{2h + 1 - j}}{2h + 1} L_{-1} \phi_{h-j+1} + \frac{j \brac{8h + 1 - 3j}}{2 \brac{2h + 1}} \phi_{h-j}} \ket{\phi}.
\end{split}
\end{equation}
Recalling \eqnref{eqnDefAStates}, we can recast this result in the form
\begin{equation}
\frac{j \brac{8h + 1 - 3j}}{2 \brac{2h + 1}} \ket{A^{\brac{j}}} = \frac{3 \brac{2h + 1 - j}}{2h + 1} L_{-1} \ket{A^{\brac{j-1}}} - \brac{L_{-2} - \frac{3}{2 \brac{2h + 1}} L_{-1}^2} \ket{A^{\brac{j-2}}}.
\end{equation}
The prefactor on the left hand side vanishes (thus $\ket{A^{\brac{j}}}$ is not determined) when $j = 0$ or $3j = 8h + 1$.  But the latter can only occur if $2p - 3 = 8h + 1 \in 3 \ZZ$, hence $p \in 3 \ZZ$.  As $p' = 3$ and $p$ must be coprime, we therefore find that for all $j > 0$,
\begin{equation} \label{eqnM3pRec}
\ket{A^{\brac{j}}} = \frac{6 \brac{2h + 1 - j}}{j \brac{8h + 1 - 3j}} L_{-1} \ket{A^{\brac{j-1}}} + \frac{3}{j \brac{8h + 1 - 3j}} L_{-1}^2 \ket{A^{\brac{j-2}}} - \frac{2 \brac{2h + 1}}{j \brac{8h + 1 - 3j}} L_{-2} \ket{A^{\brac{j-2}}}.
\end{equation}
With the initial conditions $\ket{A^{\brac{0}}} = \ket{0}$ and $\ket{A^{\brac{-1}}} = 0$, this recursively determines all the terms of the operator product expansion, \eqnref{eqnOPEGeneral}, for the $\MinMod{3}{p}$ models.  Noting that for these models $c = -2h \brac{8h-5} / \brac{2h+1}$, this relation effortlessly reproduces the results of \secref{secAlgebra} (for these models) and more:
\begin{align}
\ket{A^{\brac{1}}} &= 0, \notag \\
\ket{A^{\brac{2}}} &= -\tfrac{2h+1}{8h-5} L_{-2} \ket{0}, \notag \\
\ket{A^{\brac{3}}} &= -\tfrac{1}{2} \tfrac{2h+1}{8h-5} L_{-3} \ket{0}, \notag \\
\ket{A^{\brac{4}}} &= \tfrac{2h+1}{\brac{8h-5} \brac{8h-11}} \Bigl[ -3 \brac{h-1} L_{-4} + \tfrac{1}{2} \brac{2h+1} L_{-2}^2 \Bigr], \notag \\
\ket{A^{\brac{5}}} &= \tfrac{2h+1}{\brac{8h-5} \brac{8h-11}} \Bigl[ -2 \brac{h-1} L_{-5} + \tfrac{1}{2} \brac{2h+1} L_{-3} L_{-2} \Bigr], \\
\ket{A^{\brac{6}}} &= \tfrac{2h+1}{\brac{8h-5} \brac{8h-11} \brac{8h-17}} \Bigl[ - \brac{12h^2 - 38h + 23} L_{-6} + \brac{2h+1} \brac{3h-5} L_{-4} L_{-2} \Bigr. \notag \\*
&\mspace{305mu} \Bigl. + \brac{2h+1} \brac{h-2} L_{-3}^2 - \tfrac{1}{6} \brac{2h+1}^2 L_{-2}^3 \Bigr], \notag \\
\ket{A^{\brac{7}}} &= \tfrac{2h+1}{\brac{8h-5} \brac{8h-11} \brac{8h-17}} \Bigl[ - \tfrac{3}{4} \brac{12h^2 - 38h + 23} L_{-7} + \tfrac{1}{2} \brac{2h+1} \brac{4h-7} L_{-5} L_{-2} \Bigr. \notag \\*
&\mspace{230mu} \Bigl. + \tfrac{3}{4} \brac{2h+1} \brac{2h-3} L_{-4} L_{-3} - \tfrac{1}{4} \brac{2h+1}^2 L_{-3} L_{-2}^2 \Bigr]. \notag
\end{align}

One can similarly derive recursion relations for higher $p'$.  For example, the $\MinMod{4}{p}$ models have $h = \tfrac{1}{2} \brac{p-2}$ and singular vector
\begin{equation} \label{eqnM4pFirstSV}
\brac{L_{-3} - \frac{2}{h} L_{-2} L_{-1} + \frac{1}{h \brac{h+1}} L_{-1}^3} \ket{\phi},
\end{equation}
leading to the recursion relation
\begin{multline} \label{eqnM4pRec}
\ket{A^{\brac{j}}} = \Biggl\{ \frac{3 \brac{2h+2-j} \brac{2h+1-j} - 2 \brac{h+1} \brac{3h+1-j}}{j \brac{h-j} \brac{3h+1-j}} L_{-1} \ket{A^{\brac{j-1}}} \Biggr. \\*
+ \frac{3 \brac{2h+2-j}}{j \brac{h-j} \brac{3h+1-j}} L_{-1}^2 \ket{A^{\brac{j-2}}} - \frac{2 \brac{h+1} \brac{2h+2-j}}{j \brac{h-j} \brac{3h+1-j}} L_{-2} \ket{A^{\brac{j-2}}} \\*
+ \frac{1}{j \brac{h-j} \brac{3h+1-j}} L_{-1}^3 \ket{A^{\brac{j-3}}} - \frac{2 \brac{h+1}}{j \brac{h-j} \brac{3h+1-j}} L_{-2} L_{-1} \ket{A^{\brac{j-3}}} \\*
\Biggl. + \frac{h \brac{h+1}}{j \brac{h-j} \brac{3h+1-j}} L_{-3} \ket{A^{\brac{j-3}}} \Biggr\},
\end{multline}
with $\ket{A^{\brac{0}}} = \ket{0}$ and $\ket{A^{\brac{-1}}} = \ket{A^{\brac{-2}}} = 0$.  We note that $h = \frac{1}{2} p - 1$ implies that $h , 3h + 1 \notin \ZZ$, so this recursion relation does indeed define $\ket{A^{\brac{j}}}$ for all $j > 0$.  Whilst this recursion relation may appear somewhat intimidating, its derivation and utilisation are easy to implement on a computer algebra system.

It is very interesting to compare the $p'=3$ and $p'=4$ recursion relations in the single case where they both apply:  $\MinMod{3}{4}$.  As shown in \tabref{tabM34Rec}, their respective solutions $\ket{A^{\brac{j}}}$ agree for $j \leqslant 5$, but do \emph{not} agree for $j \geqslant 6$.  However, the difference between the two $j = 6$ solutions can be checked to be
\begin{equation}
\frac{90}{7007} \brac{L_{-6} + \frac{22}{9} L_{-4} L_{-2} - \frac{31}{36} L_{-3}^2 - \frac{16}{27} L_{-2}^3} \ket{0},
\end{equation}
a multiple of the vacuum singular vector (given in \eqnref{eqnM34VacSing}).  Likewise, higher-grade solutions differ by descendants of this singular vector. At first, it seems somewhat astonishing that these recursion relations, derived from the principal singular vectors of the simple current $\alg{Vir}$-module, may be used to compute the singular vector of corresponding vacuum module.  On second thoughts though, this is perhaps not entirely unexpected given that the solutions to the recursion relations must be consistent.  But more fundamentally, this is quite consistent with the concluding remarks of \secref{secGenSingVect}:  In the extended chiral algebra framework, all information concerning the spectrum should be derivable from the two principal singular vectors of the simple current module.

\begin{table}
\begin{center}
\setlength{\extrarowheight}{4pt}
\begin{tabular}{|C||C|C|C|}
\hline
j & \text{Basis} & p' = 3 \text{ solution} & p' = 4 \text{ solution} \\
\hline
\hline
2 & \brac{L_{-2}} & \brac{2} & \brac{2} \\
3 & \brac{L_{-3}} & \brac{1} & \brac{1} \\
4 & \brac{L_{-4}, L_{-2}^2} & \brac{\tfrac{3}{7}, \tfrac{2}{7}} & \brac{\tfrac{3}{7}, \tfrac{2}{7}} \\
5 & \brac{L_{-5}, L_{-3} L_{-2}} & \brac{\tfrac{2}{7}, \tfrac{2}{7}} & \brac{\tfrac{2}{7}, \tfrac{2}{7}} \\
6 & \brac{L_{-6}, L_{-4} L_{-2}, L_{-3}^2, L_{-2}^3} & \brac{\tfrac{2}{13}, \tfrac{2}{13}, \tfrac{6}{91}, \tfrac{4}{273}} & \brac{\tfrac{76}{539}, \tfrac{6}{49}, \tfrac{83}{1078}, \tfrac{12}{539}} \\
7 & \brac{L_{-7}, L_{-5} L_{-2}, L_{-4} L_{-3}, L_{-3} L_{-2}^2} & \brac{\tfrac{3}{26}, \tfrac{10}{91}, \tfrac{6}{91}, \tfrac{2}{91}} & \brac{\tfrac{57}{539}, \tfrac{40}{539}, \tfrac{39}{539}, \tfrac{18}{539}} \\[1mm]
\hline
\end{tabular}
\vspace{3mm}
\caption{Solutions $\ket{A^{\brac{j}}}$ to the $p' = 3$ and $p' = 4$ recursion relations when $p = 4$ and $p = 3$, respectively.  Both describe the \ope{} of the Ising model simple current (the free fermion) with itself.  The second column gives an ordered basis (when acting on $\ket{0}$) for the subspace of the vacuum Virasoro module of grade $j$, and the third and fourth columns give the coefficients of the solutions with respect to this basis.} \label{tabM34Rec}
\end{center}
\end{table}

\subsection{Recursion Relations for $\mathcal{S}$-Eigenvalues} \label{secSEigRecursive}

In \secref{secSEig}, we derived general expressions for the $\mathcal{S}$-eigenvalues of a \hws{} $\ket{\psi}$ of $\func{\alg{u}}{1}$-charge $\theta \leqslant 2$.  As the charge increases, these expressions become increasingly more difficult to compute, as evidenced by \eqnsref{eqnSEigCh=0}{eqnSEigCh=2}.  In general, we may write
\begin{gather}
1 = \bracket{\psi}{\phi_{h - \theta} \phi_{\theta - h}}{\psi} \gcreq{2 \theta + 1} \bracket{\psi}{\mathcal{S} \sum_{j=0}^{2 \theta} \binom{\theta}{2 \theta - j} A^{\brac{j}}_0}{\psi} \notag \\
\Rightarrow \qquad \bracket{\psi}{\mathcal{S}}{\psi} = \sqbrac{\sum_{j=0}^{2 \theta} \binom{\theta}{2 \theta - j} \bracket{\psi}{A^{\brac{j}}_0}{\psi}}^{-1}.
\end{gather}
The problem lies in computing the terms 
\begin{equation} 
f_j \equiv \bracket{\psi}{A^{\brac{j}}_0}{\psi}
\end{equation}
for large $j$.
But this problem is tailor-made for the recursion relations of \secref{secOPERecursive}.  A recursion relation for the $\ket{A^{\brac{j}}}$ induces a similar relation for the corresponding fields $\func{A^{\brac{j}}}{w}$, and therefore for their modes.  We illustrate this for the $\MinMod{3}{p}$ models.  \eqnref{eqnM3pRec} implies that
\begin{equation}
A^{\brac{j}}_0 = - \alpha_j \brac{j-1} A^{\brac{j-1}}_0 + \beta_j \brac{j-1} \brac{j-2} A^{\brac{j-2}}_0 + \gamma_j \sum_{r \in \ZZ} \normord{L_r A^{\brac{j-2}}_{-r}}
\end{equation}
(here $\alpha_j$, $\beta_j$ and $\gamma_j$ denote the coefficients of $L_{-1} \ket{A^{\brac{j-1}}}$, $L_{-1}^2 \ket{A^{\brac{j-2}}}$ and $L_{-2} \ket{A^{\brac{j-2}}}$ in \eqnref{eqnM3pRec}, respectively), hence that
\begin{equation} \label{eqnM3pSEigRec}
f_j = - \alpha_j \brac{j-1} f_{j-1} + \Bigl[ \beta_j \brac{j-1} \brac{j-2} + \gamma_j \brac{h_{\psi} + j-2} \Bigr] f_{j-2}.
\end{equation}
With $f_0 = 1$ and $f_{-1} = 0$, it is now a trivial task to determine the $\mathcal{S}$-eigenvalues.

For example, when $p = 17$ ($h = \tfrac{15}{4}$), the \hwss{} of charge $0, \tfrac{1}{2}, 1, \ldots , \tfrac{15}{2} = 2h$ are determined by \eqnref{eqnM3pSEigRec} to have respective eigenvalues
\begin{equation}
1, 2, \tfrac{50}{11}, \tfrac{25}{2}, \tfrac{95}{2}, 380, -4940, 12350, -12350, 4940, -380, - \tfrac{95}{2}, - \tfrac{25}{2}, - \tfrac{50}{11}, -2, \text{ and } -1.
\end{equation}
It is a strong consistency test of our formalism that these computations respect the symmetry implied by \eqnDref{eqnSComm}{eqnIdentVirHWS}, which relates the $\mathcal{S}$-eigenvalues of the states of charge $\theta$ and $2h - \theta$.  Indeed, this symmetry is by no means apparent from the recursion relations.

Observe that \eqnref{eqnM3pSEigRec} may also be used to formally obtain $\mathcal{S}$-eigenvalues for higher charge states, for example a charge $8$ \hws{} of $\MinMod{3}{17}$ would have eigenvalue $\tfrac{-17}{31}$ according to our recursion formula.  Given that $\mathcal{S}$-eigenvalues of states of negative charge must vanish (\secref{secSEig}), this appears to be at odds with the $\mathcal{S}$-eigenvalue symmetry mentioned above.  But such states cannot actually be present in the theory, as they do not respect the singular vectors in the simple current module (as we will see in \secref{secSpectrum}).  Given that the above recursion relations were originally derived from these singular vectors, it is not surprising that the ``non-physical'' solutions to these relations do not satisfy the $\mathcal{S}$-eigenvalue symmetry we might otherwise expect.

\subsection{General Applicability} \label{secConjFactor}

It should be clear that the method employed to derive the recursion relations for the $\ket{A^{\brac{j}}}$ may be applied quite generally, the only obstacle being the determination of whether the common denominator of the coefficients ever vanishes (for $j \in \ZZ_+$).  It is tempting to conjecture that this denominator never vanishes, as in \eqnDref{eqnM3pRec}{eqnM4pRec}, hence that for every $\MinMod{p'}{p}$ (with $p > p' > 2$), these recursion relations determine $\ket{A^{\brac{j}}}$ for all positive $j$.  However, it turns out that for $p' = 6$, the common denominator for the coefficients of the recursion relation obtained from the level $5$ singular vector is (up to a multiplicative constant involving $h$)
\begin{equation}
j \brac{h-j} \brac{h-1-3j} \brac{2h+1-j} \brac{10h+8-3j},
\end{equation}
which vanishes for $j = h = p-2 \in \ZZ$ and $j = 2h+1 = 2p-3 \in \ZZ$.  It follows that for these two values of $j$ (as well as $j = 0$), the $p' = 6$ recursion relation does not determine $\ket{A^{\brac{j}}}$.

Remarkably, it seems that this property that the common denominator factors nicely is a general feature of $\MinMod{p'}{p}$ models.  Based on explicit calculations for $p' \leqslant 10$, we conjecture that the common denominator of the coefficients of the recursion relation derived from the grade $p'-1$ principal singular vector is (up to an $h$-dependent multiplicative constant which can be normalised away) given by
\begin{equation} \label{eqnConjCoeff}
\prod_{k = 1}^{p'-1} \bigl[ k \brac{k - 1} p - p' \brac{j + k - 1} \bigr].
\end{equation}
The factor with $k = 1$ gives the expected vanishing when $j = 0$.  The other factors give zero (for some $j$) when $p'$ divides $k \brac{k - 1} p$ (abbreviated $p' \mid k \brac{k - 1} p$), hence when $p' \mid k \brac{k - 1}$ (for some $1 < k < p'$).  In \appref{appNumTheory} (\propref{propNumTh}), we show that such $k$ only exist when $p'$ is not a prime power.  Thus for $p' = 6 , 10 , 12 , 15 , \ldots$, the corresponding recursion relation will fail to determine $\ket{A^{\brac{j}}}$ for at least two (positive) values for $j$ (the precise number is determined in \propref{propNumTh2}).

One can analogously derive similar recursion relations from descendant singular vectors.  We shall call these ``descendant recursion relations'' to distinguish them from the ``principal recursion relations'' discussed above.  It is easy to show that these descendant recursion relations have coefficients whose common denominator is that of the principal recursion relation (conjectured to be (\ref{eqnConjCoeff})), multiplied by other $j$-dependent factors which depend on the particular descendant used in the derivation.  Hence these descendant recursion relations will also fail to determine the $\ket{A^{\brac{j}}}$ for the same set of $j$ as the principal recursion relation (and may fail for other $j$ as well).

However, there are two independent principal singular vectors in the $\alg{Vir}$-Verma module $\VirVerMod{p',p}{p'-1,1}$, so we should be able to derive two independent principal recurrence relations (we have already compared the solutions of these for $\MinMod{3}{4}$ in \tabref{tabM34Rec}).  At the level of the common denominator of the coefficients of these relations, swapping between the principal singular vectors merely amounts to swapping $p'$ and $p$.  Assuming the conjecture (\ref{eqnConjCoeff}), we may therefore ask if there are $\ket{A^{\brac{j}}}$ (with $j > 0$) which are not determined by either principal recurrence relation.  Numerical studies suggest\footnote{We used \textsc{Maple} to search for such $j$, and found that no such $j$ exist when $p' , p \leqslant 1000$.} that the answer is ``no'':  The $\ket{A^{\brac{j}}}$, and hence the \ope{} of $\func{\phi}{z}$ with itself, may always be computed from these two principal recurrence relations.  However, we have not been able to construct a proof of this claim. 

Finally, we mention an intriguing reformulation of our conjectured expression (\ref{eqnConjCoeff}) for the common denominator of the principal recurrence relation derived from the principal singular vector at grade $p' - 1$.  We can express this singular vector in the form
\begin{equation}
\ket{\chi} = \sum_{\lambda \in \mathcal{P}_{p'-1}} a_{\lambda} \ L_{-\lambda_1} L_{-\lambda_2} \cdots L_{-\lambda_{\func{\ell}{\lambda}}} \ket{\phi},
\end{equation}
where $\mathcal{P}_n$ is the set of partitions $\lambda = \brac{\lambda_1 \geqslant \lambda_2 \geqslant \cdots \geqslant \lambda_{\func{\ell}{\lambda}}}$ of $n$, and the $a_{\lambda}$ are unknown (rational) coefficients.  We can follow the derivation of the corresponding recursion relation, only keeping track of the coefficient of the term in which all Virasoro modes are removed by commutation with the $\phi$-mode.  This coefficient defines the common denominator of the recursion relation, so our conjecture for this implies that
\begin{equation}
\sum_{\lambda \in \mathcal{P}_{p'-1}} a_{\lambda} \ \prod_{k=1}^{\func{\ell}{\lambda}} \sqbrac{ \brac{\lambda_k + 1} h + \sum_{i=k+1}^{\func{\ell}{\lambda}} \lambda_i - j} = \prod_{k = 1}^{p'-1} \bigl[ k \brac{k - 1} p - p' \brac{j + k - 1} \bigr], \end{equation}
where $h = \frac{1}{4} \brac{p'-2} \brac{p-2}$ (and the $a_{\lambda}$ are appropriately normalised).

We have not tried to prove this formula, but it has been checked up to $p' = 10$ (for all $p$).  It seems amazing to us that the coefficients of the singular vector should satisfy such a nice relation --- indeed, for $p' < 6$, this relation completely determines the coefficients $a_{\lambda}$ (up to an overall normalisation).  One can of course substitute in explicit expressions \cite{BenDeg88} for the singular vector coefficients, leading to bemusing identities of considerable complexity.  We will not attempt to analyse these here, but it would be very interesting to try to understand these observations at a deeper level.

\subsection{Singular Vectors and the Spectrum} \label{secSpectrum}

As remarked at the end of \secref{secGenSingVect}, it is well-established (though difficult to prove) that the chiral spectrum of a rational conformal field theory is to a large extent controlled by the singular vectors of the vacuum module (with respect to the chiral symmetry algebra).  For example, $\MinMod{3}{4}$ has a principal singular vector $\ket{\chi}$ in its vacuum $\alg{Vir}$-module $\VirVerMod{3,4}{1,1}$ at grade $6$.  The corresponding field $\func{\chi}{z}$ is therefore null, hence its modes $\chi_n$ must map \emph{any} state of the theory into a singular state.  By computing the action of $\chi_0$ on an arbitrary \hws{}, one finds that the result is singular (in fact vanishing) if and only if the dimension of the \hws{} is $0$, $\tfrac{1}{16}$ or $\tfrac{1}{2}$.

In this way, the principal vacuum singular vector determines the chiral spectrum of the theory.  We ask the obvious question of whether this property is preserved in the extended theories:  Do the principal singular vectors of the simple current module also determine the spectrum of the (extended) theory?  We have seen in \secref{secOPERecursive} that (at least for the Ising model) these singular vectors already ``know about'' the vacuum singular vector, so we expect that the answer to our question is ``yes''.

This may indeed be demonstrated in simple cases.  We suppose that $\ket{\psi}$ is a (Virasoro) \hws{} of monodromy charge $\theta$ and conformal dimension $\Delta$.  In an $\MinMod{3}{p}$ model, the singular vector (\ref{eqnM3pFirstSV}) defines a null field $\func{\chi}{z}$ (vanishing in the extended theory) whose modes have the form
\begin{equation}
\chi_n = \sum_{r \in \ZZ} \normord{L_r \phi_{n-r}} - \frac{3 \brac{h+n} \brac{h+n+1}}{2 \brac{2h+1}} \phi_n \equiv 0.
\end{equation}
We may therefore calculate
\begin{equation}
\begin{split}
0 &= \chi_{\theta - h} \ket{\psi} = \brac{\phi_{\theta - h} L_0 + \phi_{\theta - h + 1} L_{-1} - \frac{3 \theta \brac{\theta + 1}}{2 \brac{2h+1}} \phi_{\theta - h}} \ket{\psi} \\
&= \sqbrac{\Delta + \theta - \frac{3 \theta \brac{\theta + 1}}{2 \brac{2h+1}}} \phi_{\theta - h} \ket{\psi}.
\end{split}
\end{equation}
As $\phi_{\theta - h} \ket{\psi} \neq 0$, this gives a simple relation between the charge and dimension of any \hws{} in an $\MinMod{3}{p}$ model:
\begin{equation}
\Delta = \frac{\theta \brac{3 \theta - 4h + 1}}{2 \brac{2h+1}}.
\end{equation}

Clearly, this does not completely determine the spectrum of the theory\footnote{Interestingly, the $\MinMod{3}{p}$ spectrum is completely determined by this constraint \emph{and} by requiring that the charge be an integer or half-integer between $0$ and $h$.  However, this complete determination is a peculiarity of the $\MinMod{3}{p}$ models.  In general, some of these charges will also be forbidden.}.  What is needed is a second relation between the charge and dimension, and this is provided by the second principal singular vector.  As an example, in the Ising model ($h = \tfrac{1}{2}$), this second relation is easily verified to be
\begin{equation}
\brac{2 \theta - 1} \brac{2 \Delta - \frac{\theta \brac{2 \theta + 1}}{3}} = 0.
\end{equation}
Solving the two spectrum-constraining relations then gives $\theta = 0 , \tfrac{1}{2} , 1$ and $\Delta = 0 , \tfrac{1}{16} , \tfrac{1}{2}$ (respectively), as expected.

Of course, this analysis can be repeated for other models and other singular vectors, with similar results.  We restrict ourselves to a final remark.  It is possible to show explicitly (for example with $\MinMod{5}{6}$) that the constraints derived from the principal singular vectors select all fields from the Kac table, and \emph{not} just those which contribute to the D-type modular invariant.

\section*{Acknowledgements}

We would like to thank Matthias Gaberdiel and Antun Milas for interesting correspondence during the preparation of this article, and Patrick Jacob for  useful discussions.  We also benefited greatly from the reviewer's comments regarding the modules corresponding to fixed points of the simple current, and thank them for their detailed observations.  This research is supported by NSERC.

\appendix

\section{The Decomposition of $\MinMod{3}{10}$} \label{secM3,10}

In this appendix (which is a continuation of \secref{secCFTEquivs}), we illustrate fully the process of deriving a \cft{} equivalence by constructing an isomorphism of chiral algebras.  Uniquely amongst the minimal models, $\MinMod{3}{10}$ has a simple current of conformal dimension $h = 2$.  The central charge is $c = \tfrac{-44}{5}$, so the \ope{} takes the form
\begin{equation} \label{eqnOPE3,10}
\radord{\func{\phi}{z} \func{\phi}{w}} = \mathcal{S} \sqbrac{\frac{1}{\brac{z-w}^4} - \frac{5}{11} \frac{\func{T}{w}}{\brac{z-w}^2} - \frac{5}{22} \frac{\func{\partial T}{w}}{z-w} + \ldots}.
\end{equation}
Again, $\mathcal{S}$ commutes with the $\phi_n$, and its eigenvalues on the five $\alg{A}_{3,10}$-\hwss{} $\ket{\phi_{1,s}}$ ($s = 1 , \ldots , 5$) are given by \eqnsref{eqnSEigCh=0}{eqnSEigCh=2} as
\begin{equation}
\begin{matrix}
\mathcal{S} \ket{0} = \ket{0}, \qquad \mathcal{S} \ket{\phi_{1,2}} = 2 \ket{\phi_{1,2}}, \qquad \mathcal{S} \ket{\phi_{1,3}} = \frac{11}{2} \ket{\phi_{1,3}}, \\[2mm]
\mathcal{S} \ket{\phi_{1,4}} = 44 \ket{\phi_{1,4}}, \qquad \text{and} \qquad \mathcal{S} \ket{\phi_{1,5}} = -110 \ket{\phi_{1,5}}.
\end{matrix}
\end{equation}

We consider the linear combination $a \func{T}{z} + b \func{\phi}{z}$, $a,b \in \CC$.  This is a field of conformal dimension $2$ whose \ope{} with itself is easily read off from the \opes{} (\ref{eqnOPETT}), (\ref{eqnOPEPrimary}) and (\ref{eqnOPE3,10}).  It can be verified that this linear combination again defines a Virasoro field (that is, the \emph{four} singular terms of the \ope{} have the form of those in (\ref{eqnOPETT})) if and only if
\begin{equation}
a = 1, \quad b = 0 \qquad \text{or} \qquad a = \frac{1}{2}, \quad b = \pm \ii \sqrt{\frac{11}{10 \mathcal{S}}}.
\end{equation}
We define 
\begin{equation} \label{eqnDefT+-}
\func{T^{\pm}}{z} = \frac{1}{2} \func{T}{z} \pm \ii \sqrt{\frac{11}{10 \mathcal{S}}} \func{\phi}{z},
\end{equation}
and note that these Virasoro fields $\func{T^{\pm}}{z}$ correspond to a central charge of $c_{\pm} = \tfrac{-22}{5}$ (which in turn corresponds to the minimal model $\MinMod{2}{5}$).  Furthermore,
\begin{equation}
\radord{\func{T^+}{z} \func{T^-}{w}} = \normord{\func{T^+}{w} \func{T^-}{w}} + \ldots
\end{equation}
has no singular terms, as is readily checked, so the corresponding Virasoro algebras are independent.  We have therefore constructed an isomorphism of algebras:
\begin{equation} \label{eqnIso3,10}
\alg{A}_{3,10}^{\brac{4}} \cong \uealg{\alg{Vir}_{2,5}} \otimes \uealg{\alg{Vir}_{2,5}}.
\end{equation}
The superscript $^{\brac{4}}$ reminds us that this algebra is defined by the \gcrs{} with $\gamma = 4$ (only the singular terms of the \opes{} have been used in the construction of this isomorphism).

It is important to note that the Virasoro modes
\begin{equation}
L^{\pm}_n = \frac{1}{2} L_n \pm \ii \sqrt{\frac{11}{10 \mathcal{S}}} \phi_n
\end{equation}
corresponding to the fields $\func{T^{\pm}}{z}$ are only defined to act on states of integral monodromy charge (so $n \in \ZZ$).  The representation theory corresponding to the isomorphism (\ref{eqnIso3,10}) is therefore restricted to the $\alg{A}_{3,10}$-modules headed by $\ket{0}$, $\ket{\phi_{1,3}}$ and $\ket{\phi_{1,5}}$, which build the $D$-type modular invariant of $\MinMod{3}{10}$.  It will not be possible to relate the modules headed by $\ket{\phi_{1,2}}$ and $\ket{\phi_{1,4}}$ to any module of $\alg{Vir}_{2,5}^{\otimes 2}$.  This situation is analogous to that of \cite[Sec.\ 4.2]{RidSU206}, in which the equivalence of the $D$-type $\func{\group{SU}}{2}_4$ and $A$-type (diagonal) $\func{\group{SU}}{3}_1$ Wess-Zumino-Witten models was constructed.

We can identify the nature of the \hwss{} under the $\alg{Vir}_{2,5}^{\otimes 2}$-action by computing
\begin{equation}
L^{\pm}_0 \ket{0} = 0 \qquad \text{and} \qquad L^{\pm}_0 \ket{\phi_{1,3}} = \frac{1}{2} L_0 \ket{\phi_{1,3}} = \frac{-1}{5} \ket{\phi_{1,3}},
\end{equation}
using \eqnref{eqnConfDim}.  As expected, $\ket{\phi_{1,7}}$ and $\ket{\phi} = \ket{\phi_{1,9}}$ are not even \hwss{} under this action:
\begin{gather}
L^{\pm}_1 \ket{\phi_{1,7}} = \pm \frac{\ii}{\sqrt{5}} \phi_1 \phi_{-1} \ket{\phi_{1,3}} = \pm \frac{\ii}{\sqrt{5}} \ket{\phi_{1,3}} \neq 0, \\
L^{\pm}_2 \ket{\phi} = \pm \ii \sqrt{\frac{11}{10}} \phi_2 \phi_{-2} \ket{0} = \pm \ii \sqrt{\frac{11}{10}} \ket{0} \neq 0.
\end{gather}
The analysis for $\ket{\phi_{1,5}}$ requires a little more delicacy.  Let $\ket{\widetilde{\phi}_{1,5}} = \phi_0 \ket{\phi_{1,5}}$, so that
\begin{equation}
L^{\pm}_0 \ket{\phi_{1,5}} = \brac{\frac{1}{2} L_0 \pm \frac{\ii \sqrt{-1}}{10} \phi_0} \ket{\phi_{1,5}} = \frac{-1}{10} \ket{\phi_{1,5}} \pm \frac{\ii \sqrt{-1}}{10} \ket{\widetilde{\phi}_{1,5}}.
\end{equation}
By similarly computing $L^{\pm}_0 \ket{\widetilde{\phi}_{1,5}}$, we see that the $\alg{Vir}_{2,5}^{\otimes 2}$-\hwss{} are actually $\ket{\phi_{1,5}} + \ket{\widetilde{\phi}_{1,5}}$ and $\ket{\phi_{1,5}} - \ket{\widetilde{\phi}_{1,5}}$.  Their conformal dimensions are $0$ and $\tfrac{-1}{5}$, but it is not possible to say which state has which dimension without choosing whether $\sqrt{-1}$ should be $\ii$ or $-\ii$.

To summarise, the isomorphism (\ref{eqnIso3,10}) induces the following relationship between the \hwss{} of $\alg{A}_{3,10}$ and $\alg{Vir}_{2,5}^{\otimes 2}$:
\begin{equation}
\ket{0} \longleftrightarrow \ket{0}^+ \otimes \ket{0}^-, \quad \ket{\phi_{1,3}} \longleftrightarrow \ket{\tfrac{-1}{5}}^+ \otimes \ket{\tfrac{-1}{5}}^-, \quad 
\begin{matrix}
\ket{\phi_{1,5}} + \ket{\widetilde{\phi}_{1,5}} \\[2mm]
\ket{\phi_{1,5}} - \ket{\widetilde{\phi}_{1,5}}
\end{matrix}
\longleftrightarrow
\begin{matrix}
\ket{0}^+ \otimes \ket{\tfrac{-1}{5}}^- \\[2mm]
\ket{\tfrac{-1}{5}}^+ \otimes \ket{0}^-
\end{matrix}
.
\end{equation}
Here we have denoted the $\alg{Vir}_{2,5}$-\hws{} of conformal dimension $\Delta$ by $\ket{\Delta}^{\pm}$ (including a label $^{\pm}$ to distinguish the two copies of $\alg{Vir}_{2,5}$).  As the respective definitions of \hws{} are compatible with this relationship, this implies the following isomorphisms of irreducible modules (using \correfs{corIrreducible}{corIrreducibleGen}):
\begin{align}
\VirIrrMod{2,5}{1,1} \otimes \VirIrrMod{2,5}{1,1} &\cong \ExtVerMod{3,10}{1,1} \cong \VirIrrMod{3,10}{1,1} \oplus \VirIrrMod{3,10}{1,9}, \label{IsoM25M310Mod1} \\
\VirIrrMod{2,5}{1,2} \otimes \VirIrrMod{2,5}{1,2} &\cong \ExtVerMod{3,10}{1,3} \cong \VirIrrMod{3,10}{1,3} \oplus \VirIrrMod{3,10}{1,7}, \label{IsoM25M310Mod2} \\
\brac{\VirIrrMod{2,5}{1,1} \otimes \VirIrrMod{2,5}{1,2}} \oplus \brac{\VirIrrMod{2,5}{1,2} \otimes \VirIrrMod{2,5}{1,1}} &\cong \ExtVerMod{3,10}{1,5} \cong \VirIrrMod{3,10}{1,5} \oplus \VirIrrMod{3,10}{1,5}. \notag
\intertext{Of course, this last chain of isomorphisms may be more succinctly expressed in the form}
\VirIrrMod{2,5}{1,1} \otimes \VirIrrMod{2,5}{1,2} &\cong \VirIrrMod{2,5}{1,2} \otimes \VirIrrMod{2,5}{1,1} \cong \VirIrrMod{3,10}{1,5}, \label{IsoM25M310Mod3}
\end{align}
so \eqnsref{IsoM25M310Mod1}{IsoM25M310Mod3} complete the tensor product multiplication table for the irreducible $\alg{Vir}$-modules making up $\MinMod{2}{5}$.  This table has been previously derived (though indirectly) in \cite[Prop.\ 7.2.1]{WakLec01} using results on the asymptotic growth of the characters of these irreducible $\alg{Vir}$-modules.

We conclude by studying the effect of imposing the $\gamma = 5$ \gcr{} on our construction.  In particular, we impose the (equivalent) field identification
\begin{equation}
\normord{\func{\phi}{w} \func{\phi}{w}} = \mathcal{S} \sqbrac{\frac{5}{22} \normord{\func{T}{w} \func{T}{w}} - \frac{3}{22} \func{\partial^2 T}{w}}.
\end{equation}
As $\normord{\func{\phi}{w} \func{T}{w}} = \normord{\func{T}{w} \func{\phi}{w}}$, \eqnref{eqnDefT+-} gives
\begin{equation}
\normord{\func{T^+}{w} \func{T^-}{w}} = \frac{1}{2} \normord{\func{T}{w} \func{T}{w}} - \frac{3}{20} \func{\partial^2 T}{w}.
\end{equation}
As the $\func{T^{\pm}}{w}$ generate the modes defining each side of the tensor product $\alg{Vir}_{2,5} \otimes \alg{Vir}_{2,5}$, we might expect that this product vanishes.  However, we see that it cannot, as this would require
\begin{equation}
\sqbrac{L_{-2}^2 - \frac{3}{5} L_{-4}} \ket{0} = 0,
\end{equation}
and this vector is not even singular in $\MinMod{3}{10}$.  Nevertheless, we can back-substitute $T = T^+ + T^-$ to show that
\begin{equation}
\frac{1}{2} \brac{\normord{\func{T^+}{w} \func{T^+}{w}} - \frac{3}{20} \func{\partial^2 T^+}{w}} + \frac{1}{2} \brac{\normord{\func{T^-}{w} \func{T^-}{w}} - \frac{3}{20} \func{\partial^2 T^-}{w}} = 0.
\end{equation}
Applying this to the $\alg{Vir}_{2,5}^{\otimes 2}$-vacuum $\ket{0}^+ \otimes \ket{0}^-$, we therefore recover the vanishing of the $\MinMod{2}{5}$ singular vector (in each copy):
\begin{equation}
\sqbrac{\brac{L^{\pm}_{-2}}^2 - \frac{3}{5} L^{\pm}_{-4}} \ket{0}^{\pm} = 0.
\end{equation}
It follows that $\alg{A}_{3,10}^{\brac{5}}$ and $\uealg{\alg{Vir}_{2,5}} \otimes \uealg{\alg{Vir}_{2,5}}$ are not isomorphic, because the former contains a defining relation which proves that the primitive $\MinMod{2}{5}$ vacuum singular vectors are identically zero (which cannot be proven in $\uealg{\alg{Vir}_{2,5}}$).

\section{The Vanishing of an $\MinMod{3}{5}$ Singular Vector} \label{appM35}

This appendix is devoted to outlining (by example) how one can demonstrate that the second principal singular vector of the vacuum Verma module vanishes identically.  For $\MinMod{3}{5}$, this singular vector is explicitly given (up to normalisation) by
\begin{equation}
\ket{\chi} \equiv \brac{L_{-4} - \frac{55}{57} L_{-3} L_{-1} - \frac{45}{38} L_{-2}^2 + \frac{125}{57} L_{-2} L_{-1}^2 - \frac{125}{342} L_{-1}^4} \phi_{-3/4} \ket{0}.
\end{equation}
We have established in \secref{secExamples} that the first singular vector $\ket{\omega}$ vanishes identically, hence so do all its descendants.  In particular, its descendants at grade four,
\begin{align}
\ket{\xi} &\equiv L_{-1}^2 \ket{\omega} = \brac{L_{-4} + L_{-3} L_{-1} + \frac{1}{2} L_{-2} L_{-1}^2 - \frac{3}{10} L_{-1}^4} \phi_{-3/4} \ket{0} \\
\text{and} \qquad \ket{\zeta} &\equiv L_{-2} \ket{\omega} = \brac{L_{-2}^2 - \frac{3}{5} L_{-2} L_{-1}^2} \phi_{-3/4} \ket{0},
\end{align}
must vanish identically.  Note that $\ket{\chi}$ is not (obviously) a linear combination of these descendants.

We compute:
\begin{equation}
\begin{split}
L_{-4} \ket{\phi} &= \brac{\phi_{-3/4} L_{-4} + \frac{7}{4} \phi_{-19/4}} \ket{0} \\
&\gcreq{5} \phi_{-3/4} \brac{\frac{-8}{3} \phi_{-13/4} \phi_{-3/4} - \frac{5}{3} L_{-2}^2} \ket{0} + \frac{7}{96} L_{-1}^4 \ket{\phi} \\
&= \brac{\frac{-8}{3} \phi_{-3/4} \phi_{-13/4} - \frac{5}{3} L_{-2}^2 + \frac{25}{12} L_{-2} L_{-1}^2 - \frac{241}{1152} L_{-1}^4} \ket{\phi}.
\end{split}
\end{equation}
The $\phi$-bilinear may be evaluated recursively:
\begin{gather}
\brac{\phi_{-3/4} \phi_{-13/4} - \frac{5}{2} \phi_{-7/4} \phi_{-9/4} + \frac{15}{8} \phi_{-11/4} \phi_{-5/4} - \frac{15}{384} \phi_{-19/4} \phi_{3/4}} \ket{\phi} \gcreq{-1} 0, \notag \\
\brac{\phi_{-7/4} \phi_{-9/4} - \frac{1}{2} \phi_{-11/4} \phi_{-5/4} - \frac{1}{16} \phi_{-19/4} \phi_{3/4}} \ket{\phi} \gcreq{1} 0, \notag \\
\text{and} \qquad \brac{\phi_{-11/4} \phi_{-5/4} + \frac{15}{8} \phi_{-19/4} \phi_{3/4}} \ket{\phi} \gcreq{3} \frac{5}{4} L_{-4} \ket{\phi} \notag \\
\Rightarrow \quad 
\phi_{-3/4} \phi_{-13/4} \ket{\phi} = \brac{\frac{-25}{32} L_{-4} + \frac{175}{128} \phi_{-19/4} \phi_{3/4}} \ket{\phi} = \brac{\frac{-25}{32} L_{-4} + \frac{175}{3072} L_{-1}^4} \ket{\phi}.
\end{gather}
Substituting back and tidying up, we find that
\begin{equation}
\ket{\eta} \equiv \brac{L_{-4} - \frac{20}{13} L_{-2}^2 + \frac{25}{13} L_{-2} L_{-1}^2 - \frac{1}{3} L_{-1}^4} \phi_{-3/4} \ket{0} = 0.
\end{equation}

Basic linear algebra now shows that $\ket{\chi}$, $\ket{\xi}$, $\ket{\zeta}$ and $\ket{\eta}$ are linearly dependent.  Indeed,
\begin{equation}
\ket{\chi} = \frac{2725}{1482} \ket{\xi} - \frac{55}{57} \ket{\zeta} + \frac{112}{57} \ket{\eta},
\end{equation}
which completes the proof that the second principal singular vector of the vacuum $\alg{A}_{3,5}$-Verma module vanishes identically.

\section{Two Number-Theoretic Results} \label{appNumTheory}

This appendix is devoted to the proof of two simple (related) number-theoretic results, which are used in \secref{secOPERecursive}.

\begin{proposition} \label{propNumTh}
Let $p \geqslant 3$.  Then, $p \mid k \brac{k-1}$ for some $1 < k < p$ if and only if $p$ is \emph{not} a prime power ($p \neq q^n$, $q$ prime and $n > 0$).
\end{proposition}
\begin{proof}
If $p$ is a prime power, $p \mid k \brac{k-1}$ implies $p \mid k$ or $p \mid k-1$ (since $\gcd \set{k-1, k} = 1$).  But then, $p \leqslant k$ (as $k \neq 0, 1$), so there can be no $1 < k < p$ satisfying $p \mid k \brac{k-1}$.

Suppose then that $p$ is not a prime power, so we may write $p = ab$ where $a, b \in \ZZ$, $a, b > 1$, and $\gcd \set{a, b} = 1$.  We can therefore find $a', b' \in \ZZ$ such that
\begin{equation} \label{eqnBezout}
a b' - b a' = 1.
\end{equation}
Note that this property is invariant under $\brac{a',b'} \rightarrow \brac{a' + \ell a, b' + \ell b}$, for all $\ell \in \ZZ$.  Hence we may choose $\ell$ so that
\begin{equation}
0 \leqslant a \brac{b' + \ell b} = a b' + \ell p \leqslant p-1.
\end{equation}
Let $k = a \brac{b' + \ell b}$.  Then, by \eqnref{eqnBezout},
\begin{equation} \label{eqnk-1}
k-1 = a b' - 1 + \ell a b = b a' + \ell a b = b \brac{a' + \ell a},
\end{equation}
hence
\begin{equation}
k \brac{k-1} = a \brac{b' + \ell b} b \brac{a' + \ell a} = p \brac{b' + \ell b} \brac{a' + \ell a}.
\end{equation}
We have therefore constructed $k$ satisfying $0 \leqslant k < p$ and $p \mid k \brac{k-1}$.  It remains to show that $k > 1$.

If $k = 0$, $\brac{b' + \ell b} = 0$ (as $a > 1$).  But then, substituting $b'$ into \eqnref{eqnBezout} gives $-b \brac{a' + \ell a} = 1$, contradicting $b > 1$.  If $k = 1$, we derive a similar contradiction (from \eqnref{eqnk-1}), completing the proof.
\end{proof}

\begin{proposition} \label{propNumTh2}
$p \mid k \brac{k-1}$ for precisely $2^n - 2$ integers $k$ in the range $1 < k < p$, where $n$ is the number of distinct prime divisors of $p$.
\end{proposition}
\begin{proof}
The proof of \propref{propNumTh} shows that every factoring of $p$ into coprime integers $a, b > 1$ produces a unique integer $1 < k < p$ such that $p \mid k \brac{k-1}$.  There are of course $2^n - 2$ such factorings, corresponding to the ways of partitioning the distinct prime factors of $p$ into two non-empty subsets (the missing two correspond to choosing either $a$ or $b$ to be $1$, corresponding to the empty subset).  We need to show that this map from the set of factorings into the set of $k$ satisfying the required conditions is a bijection.

We first show that different factorings give different $k$.  Let $p = ab = \alpha \beta$, $a \neq \alpha$, be two such factorings, producing $k$ and $\kappa$ respectively.  Then, there must exist a prime $q$ dividing one of $a$ and $\alpha$, but not the other (otherwise $a$ would equal $\alpha$).  Without loss of generality, suppose $q \mid a$ but $q \nmid \alpha$.  Then $q \mid p$, so $q \mid \beta$, and
\begin{equation}
k = a \brac{b' + \ell b} = 0 \pmod{q} \qquad \text{but} \qquad \kappa = \alpha \brac{\beta' + \lambda \beta} = \alpha \beta' = 1 \pmod{q},
\end{equation}
since $\alpha \beta' - \beta \alpha' = 1$.  Therefore, $k \neq \kappa$.

We now show that every $k$ between $1$ and $p$ that satisfies $p \mid k \brac{k-1}$ corresponds to such a factoring of $p$.  Given such a $k$, we set
\begin{equation}
a = \gcd \set{k, p} \qquad \text{and} \qquad b = \gcd \set{k-1, p}.
\end{equation}
Then, $\gcd \set{a,b} = 1$ because $\gcd \set{k,k-1} = 1$, and $ab = p$ since $p \mid k \brac{k-1}$.  Finally, $a = 1$ implies $p \mid k-1 > 0$, hence $p \leqslant k-1$, a contradiction as $k < p$.  $b=1$ also contradicts this, so we have the required factoring.  We complete the proof by showing that this factoring $p = ab$ does indeed produce our original integer $k$.  Clearly we can trivially satisfy $a b' - b a' = 1$ by choosing
\begin{equation}
b' = \frac{k}{a} = \frac{k}{\gcd \set{k, p}} \qquad \text{and} \qquad a' = \frac{k-1}{b} = \frac{k-1}{\gcd \set{k-1, p}}.
\end{equation}
The integer produced by the proof of \propref{propNumTh} is thus
\begin{equation}
a \brac{b' + \ell b} = k + \ell p,
\end{equation}
and clearly $\ell = 0$ as this integer must lie between $1$ and $p$.
\end{proof}


\begin{thebibliography}{10}

\bibitem{RidSU206}
P~Mathieu and D~Ridout.
\newblock {The Extended Algebra of the Minimal Models}.
\newblock {\em Nucl. Phys.}, B765:201--239, 2007.
\newblock \texttt{arXiv:hep-th/0609226}.

\bibitem{KedSum95}
R~Kedem, B~McCoy, and E~Melzer.
\newblock {The Sums of Rogers, Schur and Ramanujan and the Bose-Fermi
  Correspondence in $\left( 1+1 \right)$-Dimensional Quantum Field Theory}.
\newblock In P~Bouwknegt \emph{et al}, editor, {\em Recent Progress in
  Statistical Mechanics and Quantum Field Theory}, pages 195--219. World
  Scientific, New Jersey, 1995.
\newblock \texttt{arXiv:hep-th/9304056}.

\bibitem{FeiAnn92}
B~Feigin, T~Nakanishi, and H~Ooguri.
\newblock {The Annihilating Ideals of Minimal Models}.
\newblock {\em Int. J. Mod. Phys.}, A7:217--238, 1992.

\bibitem{LepStr85}
J~Lepowsky and M~Primc.
\newblock {\em {Structure of the Standard Modules for the Affine Lie Algebra
  $A_1^{\left( 1 \right)}$}}, volume~46 of {\em Contemporary Mathematics}.
\newblock American Mathematical Society, Providence, 1985.

\bibitem{JacQua06}
P~Jacob and P~Mathieu.
\newblock {A Quasi-Particle Description of the $M \left( 3 , p \right)$
  Models}.
\newblock {\em Nucl. Phys.}, B733:205--232, 2006.
\newblock \texttt{arXiv:hep-th/0506074}.

\bibitem{JacEmb06}
P~Jacob and P~Mathieu.
\newblock {Embedding of Bases: From the $M \left( 2 , 2 \kappa + 1 \right)$ to
  the $M \left( 3 , 4 \kappa + 2 - \delta \right)$ Models}.
\newblock {\em Phys. Lett.}, B635:350--354, 2006.
\newblock \texttt{arXiv:hep-th/0511040}.

\bibitem{FeiVer02}
B~Feigin, M~Jimbo, and T~Miwa.
\newblock {Vertex Operator Algebra Arising from the Minimal Series $M \left( 3
  , p \right)$ and Monomial Basis}.
\newblock {\em Prog. Math. Phys.}, 23:179--204, 2002.
\newblock \texttt{arXiv:math.QA/0012193}.

\bibitem{BytFer99}
A~Bytsko.
\newblock {Fermionic Representations for Characters of $M \left( 3,t \right)$,
  $M \left( 4,5 \right)$, $M \left( 5,6 \right)$, and $M \left( 6,7 \right)$
  Minimal Models and Related Rogers-Ramanujan Type and Dilogarithm Identities}.
\newblock {\em J. Phys.}, A32:8045--8058, 1999.
\newblock \texttt{arXiv:hep-th/9904059}.

\bibitem{FeiMon05}
B~Feigin, M~Jimbo, T~Miwa, E~Mukhin, and Y~Takeyama.
\newblock {A Monomial Basis for the Virasoro Minimal Series $M \left( p , p'
  \right)$: The Case $1 < p'/p < 2$}.
\newblock {\em Comm. Math. Phys.}, 257:395--423, 2005.
\newblock \texttt{arXiv:math.QA/0405468}.

\bibitem{WelFer05}
T~Welsh.
\newblock {Fermionic Expressions for Minimal Model Virasoro Characters}.
\newblock {\em Mem. Amer. Math. Soc.}, 175:1--160, 2005.
\newblock \texttt{arXiv:math.co/0212154}.

\bibitem{RidMin07}
P~Mathieu and D~Ridout.
\newblock {Module Bases and Characters for the Minimal Model Extended
  Algebras}.
\newblock (In preparation).

\bibitem{BelInf84}
A~Belavin, A~Polyakov, and A~Zamolodchikov.
\newblock {Infinite Conformal Symmetry in Two-Dimensional Quantum Field
  Theory}.
\newblock {\em Nucl. Phys.}, B241:333--380, 1984.

\bibitem{DiFCon97}
P~Di Francesco, P~Mathieu, and D~S\'{e}n\'{e}chal.
\newblock {\em {Conformal Field Theory}}.
\newblock Graduate Texts in Contemporary Physics. Springer-Verlag, New York,
  1997.

\bibitem{DotOpe85}
V~Dotsenko and V~Fateev.
\newblock {Operator Algebra of Two-Dimensional Conformal Theories with Central
  Charge $C \leqslant 1$}.
\newblock {\em Phys. Lett.}, B154:291--295, 1985.

\bibitem{ZamNon85}
A~Zamolodchikov and V~Fateev.
\newblock {Nonlocal (Parafermion) Currents in Two-Dimensional Conformal Quantum
  Field Theory and Self-Dual Critical Points in $Z_N$-Symmetrical Statistical
  Systems}.
\newblock {\em Sov. Phys. JETP}, 62:215--225, 1985.

\bibitem{ZamInf85}
A~Zamolodchikov.
\newblock {Infinite Additional Symmetries in Two-Dimensional Conformal Quantum
  Field Theory}.
\newblock {\em Theor. Math. Phys.}, 65:1205--1213, 1985.

\bibitem{BenDeg88}
L~Benoit and Y~Saint-Aubin.
\newblock {Degenerate Conformal Field Theories and Explicit Expression for Some
  Null Vectors}.
\newblock {\em Phys. Lett.}, B215:517--522, 1988.

\bibitem{SchSim90}
A~Schellekens and S~Yankielowicz.
\newblock {Simple Currents, Modular Invariants and Fixed Points}.
\newblock {\em Int. J. Mod. Phys.}, A5:2903--2952, 1990.

\bibitem{CamGra98}
J~Camino, A~Ramallo, and J~Sanchez de~Santos.
\newblock {Graded Parafermions}.
\newblock {\em Nucl. Phys.}, B530:715--741, 1998.
\newblock \texttt{arXiv:hep-th/9805160}.

\bibitem{JacGra02}
P~Jacob and P~Mathieu.
\newblock {Graded Parafermions: Standard and Quasiparticle Bases}.
\newblock {\em Nucl. Phys.}, B630:433--452, 2002.
\newblock \texttt{arXiv:hep-th/0201156}.

\bibitem{FeiCoh88}
B~Feigin and D~Fuchs.
\newblock {Cohomology of some Nilpotent Subalgebras of the Virasoro and
  Kac-Moody Lie Algebras}.
\newblock {\em J. Geom. Phys.}, 5:209--235, 1988.

\bibitem{WakLec01}
M~Wakimoto.
\newblock {\em {Lectures on Infinite-Dimensional Lie Algebra}}.
\newblock World Scientific, New Jersey, 2001.

\end{thebibliography}
\end{document}